\documentclass[preprint, 12pt]{elsarticle}
\usepackage{graphics}
\usepackage{amsmath}
\usepackage{amsthm}
\usepackage{fancyhdr}
\usepackage{color}
\usepackage{amssymb}
\usepackage{latexsym}
\usepackage{graphicx}
\usepackage{hyperref}
\hypersetup{backref, colorlinks=true}
\usepackage{verbatim}
\usepackage{hyphenat}
\usepackage{cleveref}
\usepackage{subfig}
\usepackage{hyperref}
\hypersetup{backref, colorlinks=true}
\usepackage{hyphenat}
\usepackage{cleveref}

\newcommand{\harm}{\text{h}}
\newcommand{\anh}{\text{anh}}
\newcommand{\eff}{\text{eff}}
\newcommand{\har}{\text{h}}
\newcommand{\tot}{\text{tot}}
\newcommand{\po}{\text{p}}

\newcommand{\bfk}{{\mathbf{k}}}
\newcommand{\bfP}{{\mathbf{P}}}
\newcommand{\hbfP}{{\hat{\mathbf{P}}}}
\newcommand{\hP}{\hat{P} }

\newcommand{\bfx}{{\mathbf{x}}}
\newcommand{\SU}{{\mathbb{S}}}
\newcommand{\Ham}{{\mathcal{H}}}
\newcommand{\imagI}{\imath}

\begin{document}

\begin{frontmatter}
\title{Flexoelectricity and the Entropic Force between Fluctuating Fluid Membranes}

\author[a]{Kosar Mozaffari}
\author[b]{Fatemeh Ahmadpoor}
\cortext[cor1]{Corresponding author}
\author[a,c]{Pradeep Sharma \corref{cor1}}
\ead{psharma@central.uh.edu}
\address[a]{Department of Mechanical Engineering, University of Houston, Houston, TX 77204, USA }
\address[b]{Department of Mechanical and Industrial Engineering, New Jersey Institute of Technology, Newark, NJ 07102}
\address[c]{Department of Physics, University of Houston, Houston, TX 77204, USA }

\begin{abstract}
Biological membranes undergo noticeable thermal fluctuations at physiological temperatures.
When two membranes approach each other, they hinder the out of plane fluctuations of the other. This hindrance leads to an entropic repulsive force between membranes which, in an interplay with attractive and repulsive forces due to other
sources, impacts a range of biological functions: cell adhesion, membrane fusion, self-assembly, binding-unbinding transition
among others. In this work, we take cognizance of the fact that biological membranes are not purely mechanical entities and, due to the phenomenon of
flexoelectricity, exhibit a coupling between deformation and electric polarization. The ensuing coupled mechanics-electrostatics-statistical mechanics problem is analytically intractable. We use a variational perturbation method to analyze, in closed-form, the contribution of flexoelectricity to the entropic force between two fluctuating membranes and discuss its possible physical implications. We find that flexoelectricity leads to a correction that switches from an enhanced \emph{attraction} at close membrane separations and an enhanced \emph{repulsion} when the membranes are further apart. 
\end{abstract}


\end{frontmatter}

\section{Introduction}\label{sec:Introduction}

\begin{figure}[hbt!]
    \centering
    \includegraphics[width=0.7
\linewidth]{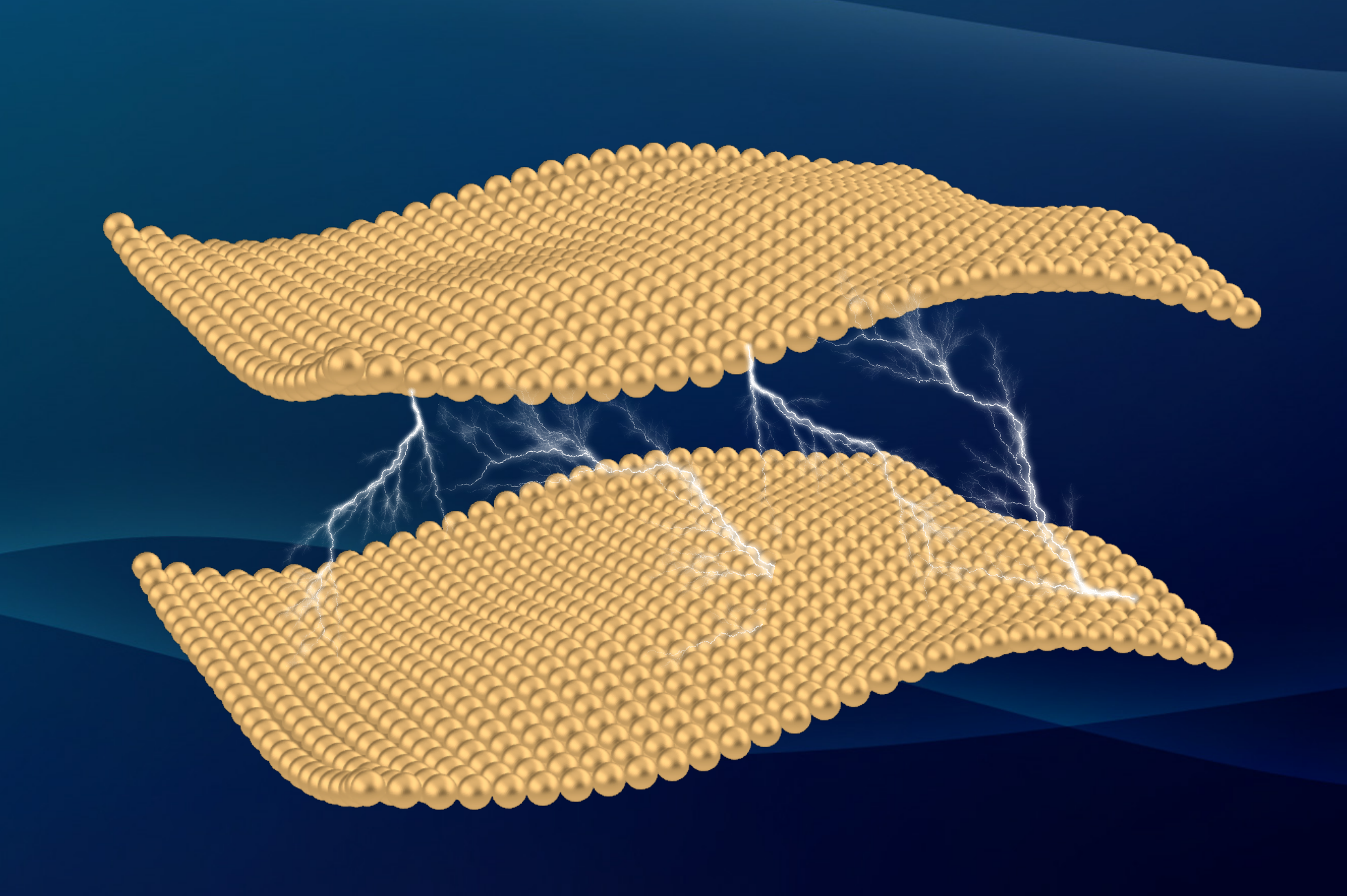}
    \caption{\small Schematic of a pair of fluctuating fluid membranes separated by a mean distance of $d$. Even neutral membranes are not purely mechanical entities and flexoelectricity is a universal phenomenon whereby changes in curvature lead to polarization. Accordingly, two fluctuating membranes will also lead to a an electrostatic interaction. }
    \label{fig:mem_flex}
\end{figure}

Consider the following biological observations: (1) a lipid bilayer wrapping around a particle or another biological entity underlying endocytosis~\cite{pastan1985pathway}, (2) two cells adhering as a precursor to fusion~\cite{fisher1993force,chernomordik1995lipids}, (3) unbinding of a cell subsequent to adhesion~\cite{lipowsky1991adhesion,lipowsky1987erratum}, and (4) formation of a stable aggregate of vesicles with a stable mean separation distance~\cite{helfrich1978steric}. These are all examples of the rather complex interplay of repulsive and attractive forces between cells and vesicles. Adhesion requires the attractive forces to be dominant. Formation of a stable aggregate, including the experimentally observed lamellar phase in vesicles, alludes to an equilibrium balance between the attractive and repulsive forces, and finally, unbinding transition is an indicator of repulsive forces dominating the interaction. The biophysical consequences are evident e.g., to name just a few; fusion is central to the beginning of life~\cite{fisher1993force}, while cell adhesion can play a critical role in bacterial infection~\cite{fisher1993force,bonazzi2011impenetrable}. \\

In the context of \emph{neutral} membranes, \emph{circa} 1977, the key forces known to mediate between biological membranes{\footnote{We will use the word biological membrane and fluid membrane interchangeably. A fluid membrane is the central component of the biological membrane but the latter is richer with several other embellishments e.g., embedded proteins. This distinction is of not much consequence to the current work. Furthermore, the word ``membrane" connotes something different within the biophysics community than for mechanicians. In the biophysics community, whose nomenclature we have adopted, a membrane is merely any two-dimensional elastic surface and can resist bending.}} were the van der Waals attraction and the so-called hydration (repulsive) force. The ubiquitous van der Waals force provides for a weak long-range attraction that varies as $1/d^3$ for close separations and scales as $1/d^6$ for larger distances~\cite{helfrich1984undulations,ninham1970van}. Here, $d$ is the mean separation distance between the membranes. In contrast, hydration force is an extremely short-range repulsion~\cite{rand1989hydration} which drops off exponentially after just a few Angstroms{\footnote{There is quite a bit of debate regarding the precise physical origins of the hydration force but that issue is not central to the current work and we refer the reader to the following reference:~\cite{lipowsky1995structure}.}}. Several observations led Helfrich~\cite{helfrich1978steric} to explore the possibility of another (un-acknowledged until then) force of a repulsive character. He argued that  in self-assembled systems, an explanation of observed equilibrium separation distance (10’s to 100 of nm) could only be explained by postulating a long-range repulsive force. He further pointed out that some vesicles didn't appear to cohere at all and subjecting the membranes to tension promoted adhesion. Helfrich's central idea to reconcile these observations can be summarized as follows: typical fluid membranes have a bending modulus  in the range of $15-25$ $k_BT$. Thus, the membranes are quite flexible and undergo noticeable mechanical undulations at physiological temperatures. Consider now two membranes at equilibrium with a thermal bath, placed parallel to each other and undergoing displacement fluctuations as shown in Figure~\ref{fig:mem_flex}. When sufficiently far apart, each will fluctuate freely un-affected by the the presence of the other.  However, when brought close together, the membranes will hinder each other's out-of-plane fluctuations. This leads to a decrease in the entropy of each membrane and hence an increase in free-energy. Closer the membranes are brought together, larger is the increase in the free-energy. This interaction therefore leads to a repulsive force that acts to push apart the membranes or alternative an external pressure is then necessary to keep the two membranes at a fixed distance apart. The entropic nature of the repulsive force is evident, and predicated on a variety of physical arguments and mathematical approximations, Helfrich~\cite{helfrich1978steric} proposed that this force scales as $1/d^3$. The entropic force, in sharp contrast to the only other known repulsive hydration force, is decidedly long-range and competes with the van der Waals force at all distances~\cite{helfrich1984undulations,ninham1970van,israelachvili1992entropic,milner1992flory,lipowsky1986unbinding,lipowsky1989binding}. We remark that, placing membranes under tension reduces fluctuations and therefore promotes adhesion (consistent with experimental observations). Furthermore, due to the competition between two long range forces (van der Waals and entropic), self-assembly is also readily explained. \\

If the change in the elastic energy of a single fluctuating membrane is described by a harmonic function, it is rather simple to analytically solve the underlying statistical mechanics problem. However, for two interacting membranes, we must impose an additional constraint: the membranes cannot interpenetrate each other. This constraint poses a formidable challenge in the ensuing statistical mechanics problem, rendering it anharmonic even for a harmonic elastic energy and an exact analytical solution becomes all but impossible. Helfrich~\cite{helfrich1978steric} first solved this problem using a simple approximation. As we will elaborate further in the next section, he simplified the inter-penetrability constraint in that rather than imposing this constraint at every single point, he assumed the membrane to obey it ``on average". Based on this simplification, he successfully obtained an approximate solution for the entropic pressure---$\alpha (k_bT)^2/\kappa_b d^3$. Here $\alpha$ is a constant and $\kappa_b$ is the bending modulus of the membrane. Since Helfrich's pioneering work, this topic has attracted significant attention in the literature spanning nearly five decades{\footnote{The literature is understandably extensive and we do not claim to be comprehensive in our citations. The papers we do mention, and the references therein however provide an adequate starting point to explore this topic.} ranging from improved analytical approximations~\cite{hanlumyuang2014revisiting,bachmann2001fluctuation,lu2015effective,janke1986fluctuation,sharma2013entropic,freund2013entropic,liang2016fluctuating,schneider1984thermal,morse1994fluctuations,morse1995statistical,michalet1994fluctuating,seifert1995concept,bachmann1999strong} (which provide a refined estimate of $\alpha$) to numerical treatments~\cite{hanlumyuang2014revisiting,janke1986fluctuation,gompper1989steric,janke1987fluctuation} of one or more fluctuating membranes. Interestingly, this topic has also inspired exploration of entropic forces in other contexts e.g.,~\cite{liang2018method,chen2015entropic,chen2017thermal}.\\

We may, at this point, consider the entropic force between two purely mechanical membranes to be largely settled. However, we must take cognizance of the fact that even neutral membranes are not quite purely mechanical entities. The coupling between strain \emph{gradients} and
polarization is the phenomenon of flexoelectricity~\cite{krichen2016flexoelectricity,tagantsev1986piezoelectricity,zubko2013flexoelectric,nguyen2013nanoscale,ahmadpoor2015flexoelectricity,mao2014insights}. \emph{All} materials (including biological membranes and even fluid membranes) exhibit the phenomenon of flexoelectricity. Specifically, in the context of membranes, flexoelectricity is simply the change in the dipole moment upon changes in the curvature. Thus, for biological membranes that bend quite easily, this phenomenon embodies quite an expedient electromechanical coupling mechanism. The microscopic underpinning of flexoelectricity for biological membranes were established by Petrov in a sequence of pioneering works--c.f.,~\cite{petrov1993flexoelectric,petrov1996flexoelectricity,petrov1998mechanosensitivity,petrov2002flexoelectricity,petrov2006electricity,petrov2007flexoelectricity} and references therein. Numerous works appear to indicate that flexoelectricity is an important electro-mechanical coupling in the context of biology and has been implicated in ion transport~\cite{petrov1993flexoelectric}, hearing mechanism~\cite{raphael2000membrane,spector2006electromechanical,breneman2009piezo,petrov1993flexoelectric,petrov2002flexoelectricity} and tether formation~\cite{brownell2010cell,glassinger2005electromechanical}. In particular, there is now considerable evidence~\cite{breneman2009hair,brownell2001micro,krichen2016flexoelectricity,raphael2000membrane,deng2019collusion} to indicate that flexoelectricity is the major mechanism behind outer hair cell electromotility, impacting cochlear amplification and sharp frequency discrimination.\\

In this work, acknowledging the universal presence of flexoelectricity in all membranes, we attempt to assess its role in modifying the entropic force between two fluctuating fluid membranes. Specifically, we employ a variational perturbative approximation to obtain a closed-form solution to the otherwise intractable problem. We remark that incorporation of flexoelectricity exacerbates further the already difficult problem of two purely mechanical fluctuating membranes. On this note, it is germane to mention that most advanced analytical treatments of the mechanical problem (without flexoelectricity) are based on some variant of the renormalization group theory which has its origins in the high-energy physics literature~\cite{amit2005field,goldenfeld2018lectures}. Kleinert and co-workers~\cite{kleinert1989gauge,kleinert2009path} introduced the so-called variational perturbation theory to handle anharmonic problems in quantum statistical mechanics and their approach offers some advantages both in terms of its efficacy as well as transparency. In particular, Bachmann et al.~\cite{bachmann1999strong} used this approximation to provide arguably the most accurate analytical solution to the purely mechanical problem underlying two interacting and fluctuating membranes (i.e., a very refined estimate of $\alpha$). To our knowledge, there is but just a single work that has attempted to elucidate the effect of flexoelectricity on the entropic force between membranes--Bivas and Petrov~\cite{bivas1981flexoelectric}.  Their approach closely parallels that of Helfrich and is unable to obtain a complete analytical solution. However they provide interesting asymptotic limits to the scaling of the entropic force at large separation distances. \\

The arrangement of this paper is as follows: In Section~\ref{sec:Preliminary Concepts}, we briefly summarize some key preliminary concepts pertaining to the statistical mechanics of membranes and present a derivation of Helfrich's original solution. In Section~\ref{sec:Theory} we formulate the complete statistical mechanics problem of two fluctuating flexoelectric membranes and then outline the solution in Section \ref{sec:VPT}. We discuss our results in Section \ref{sec:applications} and compare the effect of flexoelectricity with hydration and van der Waals forces.

\section{Preliminary Concepts}\label{sec:Preliminary Concepts}

In this section, as a prelude to our original work in subsequent sections, we provide a summary of the key concepts pertaining to the statistical mechanics problem underpinning the entropic force between two fluctuating membranes. In particular, we present the classical solution of Helfrich~\cite{helfrich1978steric} and note the various approximations. \\

Consider a thin membrane patch occupying domain of $\SU=\{(\bfx,z)|\bfx \in (0,L)^2, z=0\}$ at zero temperature with no spontaneous curvature. Further, we assume that the deformation at each membrane point is only in the out of plane i.e., in $z$ direction. The elastic energy associated with the bending of membranes up to a quadratic order (---known in the literature as the Helfrich-Canham model~\cite{Helfrich1973, Canham1970}), is

\begin{equation}\label{eq:helfric-canham-energy-repeat}
    \Ham=\int_\SU \frac{1}{2} \kappa_b H ^2+\kappa_G K,
\end{equation}
where bending modulus and Gaussian modulus, $(\kappa_b, \kappa_G)$ are material properties that parametrize the change in elastic energy with changes in the mean curvature $H$ and Gaussian curvature $K${\footnote{The elastic energy may be augmented further by incorporating surface tension or higher order effects.}}. To consider a ``large" membrane, we will assume periodicity in the in-plane directions. Further, due to the Gauss-Bonnet theorem, the contribution of the Gaussian curvature on the energy of a membrane with periodic boundary condition is merely an additive constant and thus has no bearing on the entropic force. As is frequently done in the literature, in what follows, we will employ the Monge representation i.e., each point on the surface $\SU$ is defined by its position vector $\bfx_i=(x,y)$ and a height function $h(\bf{x)}$ describing the deformation at each point that maps the undeformed flat membrane into the fluctuating deformed membrane. With this representation, the mean and Gaussian curvatures are~\cite{abbena2017modern}:
\begin{equation}\label{eq:mg_curvature}
\begin{split}
H&=\nabla \cdot \left( \frac{\nabla h(\bfx)}{\sqrt{1+|h(\bfx)|^2}}\right)\simeq \nabla^2 h(\bfx),\\
K&=\frac{{\rm{det}}(\nabla \nabla h(\bfx))}{\left( 1+|h(\bfx)|^2\right)^2}
\simeq \frac{\partial^2 h}{\partial x^2} \frac{\partial^2 h}{\partial y^2}-\left( \frac{\partial^2 h}{\partial x \partial y}\right)^2,
\end{split}
\end{equation}
where the linearized approximation\footnote{We have performed rather arduous calculations that do \emph{not} utilize this kinematic linearization however the impact of the omitted nonlinear terms are quite small and therefore, in the interest of brevity and simplicity, we restrict ourselves to the geometrically linear setting.}  is reasonable for small deviations from the flat reference state, i.e., $\nabla h(\bfx) \ll1$. \\

In the absence of a spontaneous curvature, at zero Kelvin, the minimum elastic energy of a membrane corresponds to a flat state. However, at finite temperatures, the  membranes can undulate and the probability of finding it in a specific configuration with energy $\Ham_i$ is proportional to $\exp({-\Ham_i/k_BT})$. The probability sum of all the possible configurations is given by the partition function $\mathcal{Z}$,
\begin{equation}\label{eq:partition_function}
    \mathcal{Z}=\int \exp\left( -{\Ham[h]/k_BT} \right)\mathcal{D}[h],
\end{equation}
where $\Ham[h]$ is the bending energy of the system and $\mathcal{D}[h]$ indicates that the integration is over all kinematicalky admissible deformation modes. Therefore, the probability of occurrence $\rho[h]$ of any configuration with energy $\Ham[h]$ is
\begin{equation}
    \rho[h]=\frac{1}{\mathcal{Z}} \exp\left({-\Ham[h]/k_BT}\right).
\end{equation}
Moreover, the ensemble average of any quantity may be obtained by
\begin{equation}\label{eq:ens_avg}
    \langle \Box \rangle= \frac{1}{\mathcal{Z}} \int \Box \exp\left({-\Ham[h]/k_BT} \right)\mathcal{D}[h].
\end{equation}
By having the partition function in hand, the free energy of the system is simply
\begin{equation}\label{eq:free-energy}
    \mathcal{F}=-k_BT\log(\mathcal{Z}).
\end{equation}

We now address the situation where there are two fluctuating membranes in close vicinity of each other. Here, we note that Helfrich~\cite{helfrich1978steric} showed that by performing a simple mapping, the problem of a pair of fluctuating membrane separated by a mean-distance $d$ is equivalent to a single fluctuating membrane between two hard walls separated by a distance of $2d$. The entropic pressure($\mathcal{P}$) between the two membranes that are separated by an average distance of $d$, can be obtained from the change in free energy of the entire system with respect to the separation distance~\cite{janke1986fluctuation}
\begin{equation}\label{eq:pressure}
    \mathcal{P}=-\frac{1}{A}\frac{\partial \mathcal{F}^{(m)}}{\partial (2d)},
\end{equation}
where $A$ is the area of the membrane and $\mathcal{F}^{(m)}$ is the free energy of each membrane.\\


For a free-membrane i.e., a single membrane fluctuating freely, the partition function is simply
\begin{equation}\label{eq:partition_function_single_mem}
    \mathcal{Z}=
    \int_{-\infty}^{\infty} \exp\left( - \int_\SU \frac{1}{2k_BT} \kappa_b (\nabla^2h)^2
    \right)\mathcal{D}[h].
\end{equation}

Either by the use of the equipartition theorem or by direct integration (in Fourier space, the above integral can be cast into Gaussian form) 
we can obtain a closed-form solution~\cite{ahmadpoor2017thermal}. However, if we place this membrane between two hard walls (which is equivalent to two interacting membranes), the height function of each point can only vary from $-d$ to $d$ and the partition function is modified
\begin{equation}\label{eq:partition_function_single_mem}
    \mathcal{Z}=\int_{-d}^{d} \exp\left( -{\Ham[h]/k_BT} \right)\mathcal{D}[h].
\end{equation}

The modified partition function cannot be computed analytically. Helfrich~\cite{helfrich1978steric} proposed to avoid the hard constraint and relaxed its imposition by simply restricting the mean of the square of the height function i.e., $\langle h^2\rangle < d^2$. The relaxed constraint can be imposed by a modification of the Hamiltonian as follows
\begin{equation}\label{eq:helfrich-mod}
    \Ham=\int_\SU \frac{1}{2} \kappa_b H ^2+\frac{1}{2}\alpha h^2,
\end{equation}
where the form of $\alpha$ will be determined in due course. To make further progress we transform the the Hamiltonian to the Fourier space by expanding the height function
\begin{equation}
h(\bfx)=\sum _{\bfk\in \mathbb{K}} \hat{h}(\bfk) e^{i\bfk\cdot\bfx},
\end{equation}
where $\mathbb{K}=\left\{ \bfk: \bfk=\frac{2\pi}{L}(n_1,n_2), n_1,n_2 \in \mathbb{Z}, |\bfk| \in [\frac{2\pi}{L},\frac{2 \pi}{a}] \right\}$ and $a$ is in the order of the membrane thickness. The Hamiltonian can be recast as
\begin{equation}\label{eq:helfrich-mod}
    \Ham=L^2\sum _{\bfk\in \mathbb{K}} \frac{1}{2}| \hat{h}(\bfk)| ^2 
    \left( \kappa_b  \left| \bfk\right| ^4 +\alpha \right).
\end{equation}

The partition function is obtained as
\begin{equation}\label{eq:partition_function_single_mem}
    \mathcal{Z}=\int_{-\infty}^{\infty} \exp
    \left(
     -\frac{L^2}{k_BT}\sum _{\bfk\in \mathbb{K}} \frac{1}{2}| \hat{h}(\bfk)| ^2 
    \left( \kappa_b  \left| \bfk\right| ^4 +\alpha \right) \right) \prod_{\bfk\in \mathbb{K}}d\hat{h}(\bfk)=\prod_{\bfk\in \mathbb{K}} \sqrt{\frac{2\pi k_BT}{L^2 \left( \kappa_b  \left| \bfk\right| ^4 +\alpha \right)}},
\end{equation}
and therefore the ensemble average of thermal fluctuations in Fourier and real space can be obtained from \eqref{eq:ens_avg} as:
\begin{eqnarray}
\begin{aligned}
\langle |\hat{h}{{(\mathbf{k})}}|^2\rangle&=\frac{ k_BT}{L^2 \left( \kappa_b  \left| \bfk\right| ^4 +\alpha \right)},\\
\langle {h}^2\rangle&= \frac{1}{L^2} \int_\SU \langle {h}(\bfx)^2\rangle= \sum_{\bfk\in \mathbb{K}}  \langle |\hat{h}{{(\mathbf{k})}}|^2\rangle=\frac{k_BT}{8\sqrt{\alpha \kappa_b }}=\mu d^2 < d^2, \\
\end{aligned}
\end{eqnarray}
where, in applying the soft constraint, we have assumed that $\langle {h}^2\rangle= \mu d^2$ and $\mu<1$ therefore $\alpha$ is obtained as
\begin{equation}
\alpha=\frac{1}{\kappa_b } \left (\frac{k_BT}{8 \mu d^2}\right)^2.
\end{equation}

The free energy of the system can be obtained from \eqref{eq:free-energy} as follows
\begin{eqnarray}\label{eq:free-energy_helf}
\begin{aligned}
    \mathcal{F}&=-k_BT\log(\mathcal{Z})=-k_BT  \sum_{\bfk\in \mathbb{K}}\log \left(\sqrt{\frac{2\pi k_BT}{L^2 \left( \kappa_b  \left| \bfk\right| ^4 +\alpha \right)}} \right)\\
    &=\mathcal{O}+ \frac{k_BT}{2}\sum_{\bfk\in \mathbb{K}}\log  \left( \kappa_b  \left| \bfk\right| ^4 +\frac{1}{\kappa_b } \left (\frac{k_BT}{8 \mu d^2}\right)^2
    \right)\\
    &=\frac{1}{64 \kappa_b \mu  }\left( \frac{k_BTL}{ d} \right)^2,
    \end{aligned}
\end{eqnarray}
where $\mathcal{O}$ is an inconsequential constant. From \eqref{eq:pressure}, the entropic pressure is obtained as
\begin{equation}\label{eq:pressure_helf}
    \mathcal{P}=-\frac{1}{A}\frac{\partial \mathcal{F}}{\partial (2d)}=\frac{(k_BT)^2}{64 \kappa_b  \mu } \ \frac{1}{ d^3}.
\end{equation}

Helfrich~\cite{helfrich1984undulations,helfrich1978steric} approximated $\mu=1/6$ by assuming the average of two limiting behavior: (1) restricting only few number of points of the membrane between $-d$ to $d$ and (2) restricting every point of the membrane but allowing only one specific mode of deformation. As discussed in Section \ref{sec:Introduction}, the expression for entropic pressure shows a scaling of $1/d^3$.

\section{Formulation of the Statistical Mechanics Problem for Fluctuating Flexoelectric Membranes}\label{sec:Theory}

We now proceed to formulate the electrostatics-mechanics-flexoelectric problem for two fluctuating and interacting membranes. Consider two identical planar thin polarizable membranes of thickness $t$ whose mid-surfaces occupy domains of $\SU_1=\{(\bfx,z)|\bfx \in (0,L)^2, z=0\}$ and  $\SU_2=\{(\bfx,z)|\bfx \in (0,L)^2, z=d\}$ at zero temperature without spontaneous curvature. As before, we assume periodicity within the plane. Similar to the previous section, each point on the surface $\SU_i$ is defined by its position vector $\bfx=(x,y)$ and a height function $h_i(\bf{x)}$ which is the out of plane deformation. The state of the membrane can be described by the state variables $(\bfP,h)$ where $\bfP=(P_x,P_y,P_z)$ is the electric polarization of the membrane per unit area. A suitable quadratic form for the energy of fluid membranes incorporating flexoelectric coupling{\footnote{Some very interesting work involve derivation of the precise form of elastic energy membranes from three-dimensional theories of elasticity c.f.,~\cite{steigmann2009concise,roohbakhshan2016projection,steigmann2013well,barham2012magnetoelasticity,steigmann2018mechanics,edmiston2011analysis,ogden2011mechanics,deseri2008derivation} however we directly propose a linearized constitutive response known in the literature, and physical considerations.} can be written as~\cite{mohammadi2014theory,liu2013flexoelectricity}
\begin{equation}
    \begin{split}\label{eq:bending}
\Ham[\bfP,h]&=\sum_{i=1}^2 \left[ \int _{\SU_i}\frac{1}{2}a_i\left|\bfP_i\right|^2+f_i\bfP_i \cdot \mathbf{n}_i{H_i}+\frac{1}{2}\kappa_b^{(i)} \left(H_i\right)^2 \right]+ 
 \frac{\epsilon_0}{2}\int_{\mathbb{R}^3} |\mathbf{E}|^2,
\end{split}
\end{equation}
where the superscript $(i)$ and subscript $i$ denote the $i$th membrane and $\mathbf{n}_i(\bfx)$ stands for the normal vector to the membrane $i$ at point $\bfx$. The term $f_i$ is the flexoelectric constant, $a_i= \frac{1}{(\epsilon_i - \epsilon_0)t}$\footnote{Here, for simplicity, we take the assumption that the in plane and out of plane dielectric permittivity of the membrane are equal. The reason behind this assumption is that the in plane polarization has no effect on the entropic pressure and will be eliminated later.} and $\mathbf{E}{ = - \nabla \left(\xi_1+\xi_2 \right)}$ is the electric field. The last term in \eqref{eq:bending} is rather difficult to handle and is the nonlocal electrostatic energy of the electric field induced by polarization of the membranes. In \eqref{eq:bending}, the electric field $\mathbf{E}$ and the polarization $\bfP$ must satisfy the Maxwell's equations of electrostatics\footnote{ { The reader is referred to the discussion in~\cite{liu2013flexoelectricity} for details on obtaining the nonlocal electric field.}}:
\begin{equation} \label{eq:maxwell}
\begin{cases}
\begin{aligned}
&\text{div}[-\epsilon_{0}  \nabla \xi_i (\bfx,z)+\frac{1}{t} \bfP_i(\bfx) ]=0,\\
&{  \nabla \xi_i (\bfx,z) \cdot \bold{e}_z \to {0} \quad \text{as}\quad  |z| \to \infty.}
\end{aligned}
\end{cases}
\end{equation}

{ In the plane of the membrane, the electric field satisfies periodic boundary conditions\footnote{{ The reader is also referred to the Appendix in~\cite{grasinger2020statistical} for a discussion on constant electric field ensemble.}}}.To impose the non-interpenetration constraint, we borrow a cue from the work by Kleinert~~\cite{kleinert1999fluctuation} and Bachmann~~\cite{bachmann1999strong}, and incorporate this steric constraint by add a penalizing potential to the Hamiltonian
\begin{equation}\label{eq:Uanh}
V_{\text{p}}=\frac{1}{2} m^4\frac{d^2}{\pi^2} \tan ^2\left[\frac{\pi }{2d}\left(h_1(\bfx)- h_2(\bfx)\right) \right].
\end{equation}

The penalty potential is merely a generalized version of the soft constraint imposed by Helfrich in his original work however, we note that in the limit of $m \rightarrow 0$, we obtain the exact hard-wall constraint. Figure \ref{fig:penalty-potential} shows the penalty potential versus the normalized distance between two membranes at a specific point $\bfx$ i.e., $h(\bfx)/d=(h_1(\bfx)- h_2(\bfx))/d$ and its limiting behavior as it approaches the exact rigid-wall constraint. \\
\begin{figure}[hbt!]
    \centering
    \includegraphics[width=0.6
\linewidth]{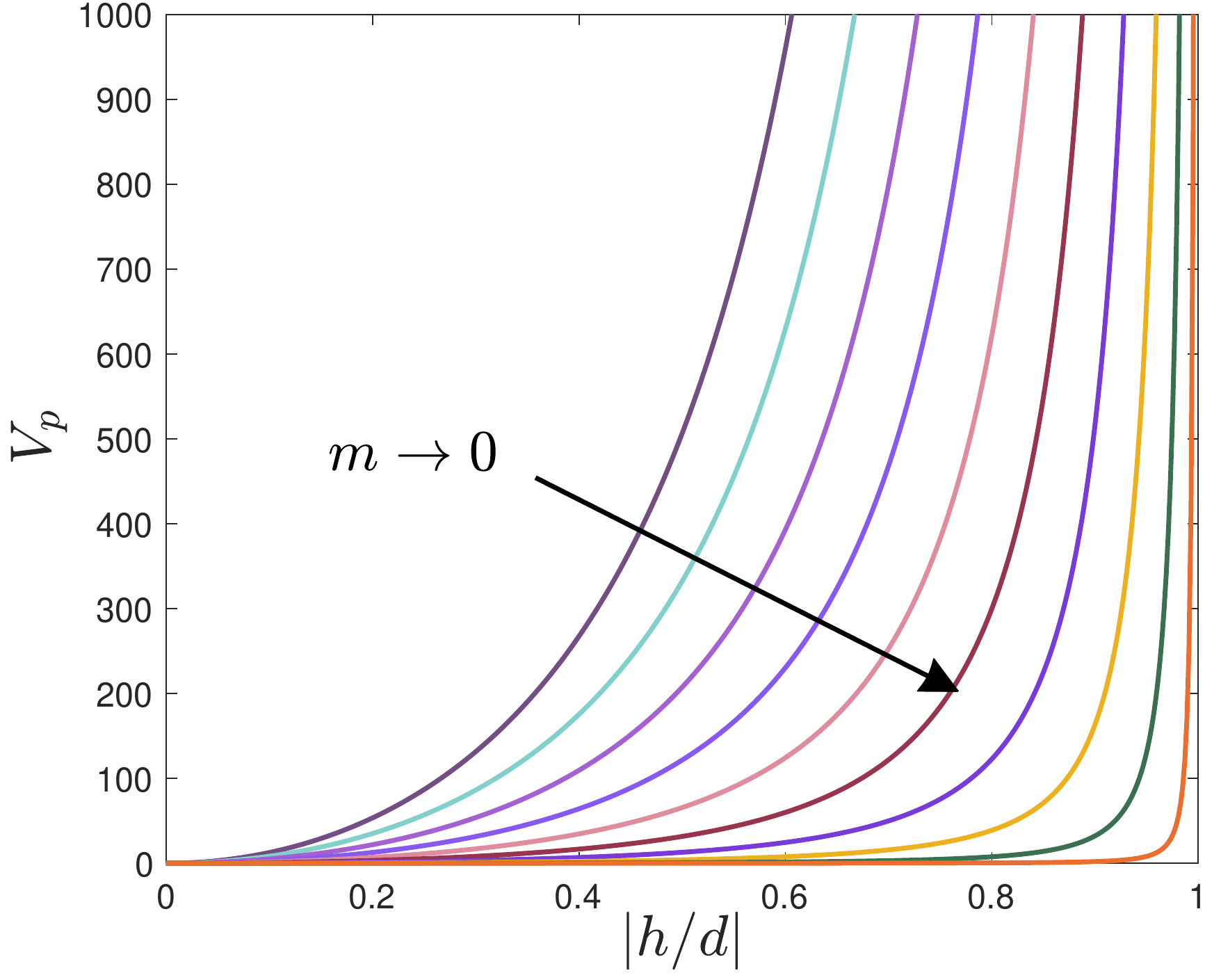}
    \caption{\small In this figure, we illustrate the limiting behavior fo the penalty potential against the normalized distance between two membranes at a specific point for different values of $m$. The exact rigid-wall constraint is obtained in the limit of $m \rightarrow 0$.}
    \label{fig:penalty-potential}
\end{figure}

The mean curvature of the membrane $i$ can be simplified to $H_i=\nabla^2 h_i(\bfx)$ in the linearized setting. For simplifying the Hamiltonian in Equation~\eqref{eq:bending} further, we introduce $u$ and $v$ such that:
\begin{equation} \label{eq-3}
\begin{aligned}
u(\bfx)=\frac{h_1(\bfx)+h_2(\bfx)}{2} \qquad , \qquad v(\bfx)=\frac{h_1(\bfx)-h_2(\bfx)}{2}.
\end{aligned}
\end{equation}

Therefore: $h_1(\bfx)=u(\bfx)+v(\bfx)$ and $h_2(\bfx)=u(\bfx)-v(\bfx)$. Since $u(\bfx)$ is the rigid body motion of the whole system, it makes no contribution to the total elastic energy, and we eliminate it in subsequent calculations. Then, the Hamiltonian can be expressed as
\begin{equation} \label{eq:Hamilt_mod}
\begin{aligned}
\Ham[\bfP,v]=&\int _{\SU}\frac{1}{2}a_1\left|\bfP_1\right|^2+\frac{1}{2}a_2\left|\bfP_2\right|^2+f_1\bfP_1 \cdot \mathbf{n}_1\left(\nabla ^2v\right)+f_2\bfP_2 \cdot \mathbf{n}_2\left(-\nabla^2v\right)+\kappa _b\left(\nabla ^2v\right)^2\\
&+\frac{1}{2}m^4\frac{d^2}{\pi ^2} \tan ^2\left(\frac{\pi }{d} v\right)
+  \frac{\epsilon_0}{2}\int_{\mathbb{R}^3} |\mathbf{E}|^2
,
\end{aligned}
\end{equation}
where for simplicity we have assumed $\SU=\SU_1 \cup \SU_2$ and $\kappa_b^{(1)}=\kappa_b^{(2)}=:\kappa_b $. 
The penalty term $V_{\text{p}}$ can be decompossed in terms of a harmonic and an anharmonic part
\begin{equation} \label{eq-6}
\begin{aligned}
V_{\text{p}}=\frac{1}{2}m^4\frac{d^2}{\pi ^2} \tan ^2\left(\frac{\pi }{d} v\right)=\frac{1}{2}\left(m^{4 }v^2+\frac{\pi ^2}{d^2}V_{\anh}(v)\right).
\end{aligned}
\end{equation} 

The anharmonic part can be expanded in Taylor series as
\begin{equation}\label{eq:int_ham}
    V_{\anh}(v)=m^4\{\alpha_4\left(v \right)^4+\alpha_6\left(\frac{\pi }{d}\right)^2\left(v \right)^6+\alpha _8\left(\frac{\pi }{d}\right)^4\left(v\right)^8+\text{...}\},
\end{equation}
where $\alpha_4=\frac{2}{3}$, $\alpha_6=\frac{17}{45}$ and $\alpha_8=\frac{62}{315}$. \\


The last term in Equation~\eqref{eq:bending} which is the nonlocal electrostatic energy of the electric field induced by polarization of the membranes, can be simplified further as~\cite{liu2013energy}
\begin{equation} \label{eq:field-engry}
\begin{aligned}
\frac{\epsilon_0}{2}\int_{\mathbb{R}^3} |\mathbf{E}|^2
=&\frac{\epsilon_0}{2}\int_{\mathbb{R}^3} |\nabla \xi_1+\nabla \xi_2 |^2\\
=&\frac{1}{2t}\int _{\SU}\int _{-t/2}^{t/2}\bfP_1 \cdot\nabla \xi _1 +2\bfP_1 \cdot\nabla \xi _2\mathrm{d}z\mathrm{d}\bfx
+\frac{1}{2t}\int _{\SU}\int _{d-t/2}^{d+t/2}\bfP_2 \cdot\nabla \xi _2
\mathrm{d}z\mathrm{d}\bfx
,
\end{aligned}
\end{equation}
where $\mathbf{E}_i=-\nabla\xi_i$ denotes the electric field induced from the polarized membrane $i$ and by definition $\xi_i$ is called the electric potential of membrane $i$. Basically, this expression is sum of two different contributions: (1) a contribution originated from the interaction of polarization of one membrane with its induced electric field and (2) the other contribution comes from the interaction of the polarization of a membrane with the electric field of the other membrane.\\

In Fourier space($\nu_i:=z-d(i-1)$):
\begin{eqnarray} \label{eq-8}
\begin{aligned}
\bfP_i(\bfx)&=\sum _{\bfk\in \mathbb{K}} \mathbf{\hat{P}}_i(\bfk)e^{i\bfk\cdot\bfx},  \\
\xi _i(\bfx,z)&=\sum _{\bfk\in \mathbb{K}} \hat{\xi}_i(\bfk,\nu_i) e^{i\bfk\cdot\bfx},
\end{aligned}
\end{eqnarray} 
where, $\mathbb{K}=\left\{ \bfk: \bfk=\frac{2\pi}{L}(n_1,n_2), n_1,n_2 \in \mathbb{Z}, |\bfk| \in [\frac{2\pi}{L},\frac{2 \pi}{a}] \right\}$.
 Substituting the transformed polarization and potential into Maxwell equations in \eqref{eq:maxwell} and solving for the electric potential $\hat{\xi}_i $, we obtain~\cite{liu2013flexoelectricity}
{\small
\begin{eqnarray} \label{eq:xi-sol}
\hat{\xi}_i(\bfk,\nu_i)=
\begin{cases}
\begin{aligned}
 &\frac{\left(\mathbf{\hat{P}}_i\right)_z}{2 t \epsilon_{0}  \left| \bfk\right|} \left(e^{\left| \bfk\right| t/2}-e^{-\left| \bfk\right| t/2}\right)e^{-\left| \bfk\right|\nu_i}
 -\frac{\imagI \hbfP_i\cdot\bfk}{2t\epsilon_0 |\bfk|^2}\left( e^{|\bfk|t/2}+ e^{-|\bfk|t/2}\right)e^{-|\bfk|\nu_i} 
 & \nu_i\geqslant t/2, \\
 &\frac{\left(\mathbf{\hat{P}}_i\right)_z}{2 t \epsilon_{0}  \left| \bfk\right| }\left(e^{\left| \bfk\right| \nu_i}-e^{-\left| \bfk\right| \nu_i}\right)e^{-\left| \bfk\right|t/2}
 -\frac{\imagI \hbfP_i\cdot\bfk}{t\epsilon_0 |\bfk|^2}\left[1-\frac{e^{-|\bfk|t/2}}{2}\left( e^{|\bfk|\nu_i}+ e^{-|\bfk|\nu_i}\right)\right] 
&   |\nu_i| \leqslant t/2, \\
 &\frac{-\left(\mathbf{\hat{P}}_i\right)_z}{2 t \epsilon_{0}  \left| \bfk\right| }\left(e^{\left| \bfk\right| t/2}-e^{-\left| \bfk\right| t/2}\right)e^{\left| \bfk\right|\nu_i} 
  -\frac{\imagI \hbfP_i\cdot\bfk}{2t\epsilon_0 |\bfk|^2}\left( e^{|\bfk|t/2}+ e^{-|\bfk|t/2}\right)e^{|\bfk|\nu_i} 
 & \nu_i\leqslant -t/2 ,
 \end{aligned}
\end{cases}
\end{eqnarray} }
where $\imagI$ is the unit imaginary number. By Equation~\eqref{eq:xi-sol}, assuming $t|\bfk| \ll 1$ which is equivalent to assuming that the wavelengths of the fluctuations are much larger than the thickness of the membrane, the first term in Equation~\eqref{eq:field-engry} can be simplified as follows
{\small
\begin{eqnarray}\label{eq:filed-energy-1}
\begin{aligned}
\int _{\SU_1}\int _{-t/2}^{t/2}\frac{1}{2t}\bfP_1 \cdot\nabla \xi _1\mathrm{d}z\mathrm{d}{\bfx} &=\sum _{\bfk\in \mathbb{K}} \frac{L^2}{2t}\int _{-t/2}^{t/2}\left[\left(\hbfP_{1}(-\bfk)\right)_z\frac{d\hat{\xi}_{ 1}(\bfk,\nu_1)}{d\nu_1}+i\hbfP_{1}(-\bfk) \cdot \bfk\hat{\xi}_1
\right]\mathrm{d}\nu_1\\
%
%
%
&=\sum _{\bfk\in \mathbb{K}} \frac{L^2}{2t \epsilon_{0} }
| \hP _{1}(\bfk)| ^2 
+O(t|\bfk|)
,
\end{aligned}
\end{eqnarray} 
}
and we have assumed $\left(\hbfP_{i}(\bfk) \right)_z=\hP _{i}(\bfk)$. Since the in-plane polarization of the membranes do not have any correlation with the out-of-plane deformation of membranes, we may neglect the in-plane polarization in the subsequent calculations. By retaining only the leading terms in Equation~\eqref{eq:field-engry} we obtain
\begin{eqnarray}\label{eq:filed-energy-1-simplified-2}
\begin{aligned}
\frac{1}{2}\int _{\SU_1}\int _{-t/2}^{t/2}\frac{1}{t}\bfP_1 \cdot\nabla \xi _1\mathrm{d}z\mathrm{d}\bfx &=\sum _{\bfk\in \mathbb{K}} \frac{L^2}{2t \epsilon_{0} }
| \hP _{1}(\bfk)|^2,\\
\frac{1}{2}\int _{\SU_2}\int _{d-t/2}^{d+t/2}\frac{1}{t}\bfP_2 \cdot\nabla \xi _2\mathrm{d}z\mathrm{d}\bfx&=\sum _{\bfk\in \mathbb{K}} \frac{L^2}{2t \epsilon_{0} }
| \hP _{2}(\bfk)| ^2 .
\end{aligned}
\end{eqnarray}

For obtaining the last integral contributing to the interaction of the electric fields in Equation~\eqref{eq:field-engry}, we use the electric potential from Equation~\eqref{eq:xi-sol}
\begin{eqnarray} \label{eq-13}
\begin{aligned}
\int _{\SU_1}\int _{-t/2}^{t/2}\frac{1}{t}\bfP_1 \cdot\nabla \xi _2\mathrm{d}z\mathrm{d}\bfx&=\sum _{\bfk\in \mathbb{K}} \frac{L^2}{t}\int _{-t/2}^{t/2}\left[\left(\hbfP_{1}(-\bfk)\right)_z
\frac{d\hat{\xi}_{ 2}(\bfk,\nu_2)}{d\nu_2}
+i\hbfP_{1}(-\bfk) \cdot \bfk\hat{\xi}_2
\right]\mathrm{d}z\\
&=\sum _{\bfk\in \mathbb{K}} \frac{-L^2e^{-\left| \bfk\right| d}\left| \bfk\right| }{2\epsilon_{0} }\left(\hbfP_{1}(-\bfk)\right)_z
\left(\hbfP_{2}(\bfk)\right)_z
+O(t|\bfk|).
\end{aligned}
\end{eqnarray} 

Expanding the height function in Fourier space we obtain
\begin{eqnarray} \label{eq:transform-vx}
\begin{aligned}
v(\bfx)&=\sum _{\bfk\in \mathbb{K}} \hat{v}(\bfk) e^{i\bfk\cdot\bfx},
\end{aligned}
\end{eqnarray} 
where $\hat{v}(\bfk)$ is the height function in Fourier space and can be obtained as follows
\begin{eqnarray} \label{eq:transform-vx}
\begin{aligned}
v(\bfk)&=\frac{1}{L^2}\int _\SU \hat{v}(\bfx) e^{-i\bfk\cdot\bfx}.
\end{aligned}
\end{eqnarray}

Therefore, assuming $a_1=a_2=:a$, the Hamiltonian in Fourier space becomes
\begin{eqnarray} \label{eq:hamiltonian-fourier}
\begin{aligned}
\Ham =&L^2\sum _{\bfk\in \mathbb{K}} \kappa_b  \left| \bfk\right| ^4\left| \hat{v}(\bfk)\right| ^2
+\frac{1}{2}a_z \hP _{1}^2(\bfk)
+\frac{1}{2}a_z \hP _{2}^2(\bfk) 
-\left| \bfk\right| ^2f_1 \hP _{1}(\bfk)\hat{v}(-\bfk)\\
&+\left| \bfk\right| ^2f_2 \hP _{2}(\bfk)\hat{v}(-\bfk)
-\frac{e^{-\left| \bfk\right| d}\left| \bfk\right| }{2 \epsilon_{0} }\hP _1(-\bfk)\hP _{2}(\bfk)
+ \frac{1}{2} m^4\left| \hat{v}(\bfk)\right| ^2
+\int_{\SU}\frac{\pi ^2}{2d^2}V_{\anh}(v(\bfx) )\\
%
\end{aligned}
\end{eqnarray}  
\\
where 
$a_z=a+\frac{1}{t \epsilon_{0} }$. The partition function is
\begin{equation} \label{eq-15}
\begin{aligned}
\mathcal{Z}&=\int e^{-\beta  \Ham[P _{1},P _{2},v]} \, \mathcal{D}[P _{1},P _{2},v]
\end{aligned}
\end{equation}

Since we do not have any anharmonicity with respect to the polarization, we can integrate it out of the Hamiltonian and obtain the modified partition function as
\begin{equation} \label{eq-15}
\begin{aligned}
\mathcal{Z}&=\int \left \{ e^{-\beta  \Ham[P _{1},P _{2},v]} \, \mathcal{D}[P _{1},P _{2},v] \, \, \mathcal{D}[P _{1},P _{2}]\right \} \mathcal{D}[v]\\
&=\mathcal{Z}_\po \times \mathcal{Z}_v,
\end{aligned}
\end{equation}  
where 
\begin{eqnarray} \label{eq:c_a}
\begin{aligned}
\mathcal{Z}_\po&=\prod _{\bfk \in \mathbb{K}}  \frac{4 \pi   }{\sqrt{a_z\text{$\beta $L}^2} \sqrt{4a_z \text{$\beta $L}^2-\frac{e^{-2\left| \bfk\right| d}\left| \bfk\right| ^2 \text{$\beta $L}^2}{a_z \epsilon_{0} ^2}}},\\
\mathcal{Z}_v&=\int e^ {-\beta L^2\sum _{\bfk\in \mathbb{K}}\left[ \frac{  \left| {\bfk}\right| ^4 \left(-2 e^{d \left| {\bfk}\right| } \epsilon_{0}  \left(f^2-a_z\kappa_b \right)+\kappa_b  \left| {\bfk}\right| \right)}{2 a_z
e^{d \left| {\bfk}\right| } \epsilon_{0} +\left| {\bfk}\right| } +\frac{1}{2}m^4\right]
\left| \hat{v}(\bfk)\right| ^2
-\beta \int _{\SU}\frac{\pi ^2}{2d^2}V_{\anh}(v)}\prod_{\bfk\in \mathbb{K}}d\hat{v}(\bfk)
\\
&=\int e^{-\beta\Ham_{\eff}} \prod_{\bfk\in \mathbb{K}}d\hat{v}(\bfk).
 \end{aligned}
\end{eqnarray}
\\

By this simplification, the effective Hamiltonian $\Ham_{\eff}$ is only the function of $\hat{v}(\bfk)$ and the polarization terms are integrated out to $\mathcal{Z}_\po$.
%
%
%

\section{Variational Perturbation Solution}\label{sec:VPT}

The partition function in $\eqref{eq:c_a}_2$ is not analytically tractable due to the penalty used to impose the steric constraint and the well-known equipartition theorem often used for harmonic Hamiltonians does not apply. Therefore, in this section, we proceed to adopt a variational perturbation theory to approximate the partition function. In partition functions of such types, a purely perturbative approach leads to a divergent expansion~\cite{kleinert1989gauge} and the specific variational approach that we use in conjunction has been shown to yield highly convergent and accurate approximations~\cite{kleinert1989gauge,bachmann1999strong}. The method consists of addition and subtraction of an unknown harmonic potential $\mathcal{V}_\har$ as follows
 \begin{eqnarray}\label{eq:vpt_add}
  \Ham_{\eff}= \Ham_{\eff}+\mathcal{V}_\har-\mathcal{V}_\har.
 \end{eqnarray}

Splitting the Hamiltonian to a harmonic and an anharmonic part and assuming the unknown harmonic potential to be $\mathcal{V}_\har=L^2\frac{1}{2} M^4\left| \hat{v}(\bfk)\right|$, we can write
 \begin{eqnarray}\label{eq:h_tr}
 \begin{aligned}
    \Ham_\har&= \Ham_{\har}+\mathcal{V}_\har-\mathcal{V}_\har\\
    &= \Ham_{\har}+L^2\frac{1}{2} M^4\left| \hat{v}(\bfk)\right| ^2-L^2\frac{1}{2} M^4\left| \hat{v}(\bfk)\right| ^2\\
   & =L^2\sum _{\bfk\in \mathbb{K}}\left[ \frac{  \left| {\bfk}\right| ^4 \left(-2 e^{d \left| {\bfk}\right| } \epsilon_{0}  \left(f^2-a_z\kappa_b \right)+\kappa_b  \left| {\bfk}\right| \right)}{2 a_z
e^{d \left| {\bfk}\right| } \epsilon_{0} +\left| {\bfk}\right| } +\frac{1}{2}\left( \sqrt{M^4-gr}
\right)^2\right]
\left| \hat{v}(\bfk)\right| ^2  ,
 \end{aligned}
 \end{eqnarray}
where we used the trial identity{\footnote{ Here we follow the \emph{square root substitution method} introduced by Kleinert~~\cite{kleinert2009path} which is a shortcut to the regular variational perturbation method~\cite{duttmann2009variational}.  }} $m^2=\sqrt{M^4-gr}$ with $r=\frac{M^4-m^4}{g }$ and $g=\frac{\pi^2}{d^2}$. From \eqref{eq:c_a}, the anharmonic part of the effective Hamiltonian becomes
 \begin{eqnarray}\label{eq:h_int}
 \begin{aligned}
    \Ham_{\anh}&=\Ham_{\eff}-\Ham_\harm=\frac{g}{2} \int _{\SU}V_{\anh}(v)\\
    &=\frac{g}{2}m^2\int _{\SU}\alpha_4{v} ^4
    +g\alpha_6{v}^6
    +\cdots
 \end{aligned}
 \end{eqnarray}

In the absence of the anharmonic Hamiltonian $\Ham_{\anh}$, the partition function $\mathcal{Z}_{\harm}$ and free energy $\mathcal{F}_{\harm}$ can be obtained analytically. The effect of the anharmonic term on the total free energy of the system $\mathcal{F}$, can be then estimated by a perturbative expansion of the free energy around harmonic free energy $\mathcal{F}_{\harm}$. Expanding the partition function of the system $Z$ around the harmonic Hamiltonian yields
\begin{eqnarray} \label{eq:partition_fn}
\mathcal{Z}=\mathcal{Z}_\po \int  \exp(-\beta(\Ham_\harm+\Ham_{\anh}))\prod_{\bfk\in \mathbb{K}}d\hat{v}(\bfk)=\mathcal{Z}_\po \mathcal{Z}_{\harm}\langle \exp(-\beta \Ham_{\anh})\rangle_{\Ham_\harm},
\end{eqnarray}
where $\langle \cdot\rangle_{\Ham_\harm}$ stands for the ensemble average with respect to the harmonic Hamiltonian $\Ham_\harm$ and the harmonic partition function $\mathcal{Z}_{\harm}$ can be obtained trivially by usual Gaussian integral relation
\begin{equation} \label{eq-20}
\begin{aligned}
\mathcal{Z}_{\harm}=\int e^{-\beta \Ham_\har } \, \prod_{\bfk\in \mathbb{K}}d\hat{v}(\bfk)=\prod _{\bfk \in \mathbb{K}} \sqrt{\frac{\pi }{\beta L^2  \left(\frac{ \left| \bfk\right| ^4 \left(-2 e^{d \left| \bfk\right| } \epsilon_{0} \left(f^2-a_z\kappa_b \right)+\kappa_b  \left| \bfk\right| \right)}{2 a_ze^{d \left| \bfk\right| } \epsilon_{0} +\left| \bfk\right| }+\frac{1}{2}m^4 \right)}}.
 \end{aligned}
\end{equation}



The free energy of the system is then obtained as
\begin{align}\label{eq:Finf}
\mathcal{F}=\mathcal{F}_\po+\mathcal{F}_v=\mathcal{F}_\po+\mathcal{F}_{\harm}-\frac{1}{\beta}\sum_{n=1}^\infty\frac{(-\beta)^n}{n!}\langle ( \Ham_{\anh})^n\rangle^c_{\Ham_\harm},
\end{align}
where $\mathcal{F}_i=-\frac{1}{\beta}\log \left(\mathcal{Z}_i\right), i=\po,v$. The superscript $\langle \cdot\rangle^c$ represents the cumulant averages\footnote{The cumulant averages, up to third order, are:
\begin{align*}
\langle \Ham_{\anh}\rangle^c_{\Ham_\harm}&=\langle \Ham_{\anh}\rangle_{\Ham_\harm}\nonumber,\\
\langle \Ham_{\anh}^2\rangle^c_{\Ham_\harm}&=\langle \Ham_{\anh}^2\rangle_{\Ham_\harm}-\langle \Ham_{\anh}\rangle^2_{\Ham_\harm}\nonumber,\\
\langle \Ham_{\anh}^3\rangle^c_{\Ham_\harm}&=\langle \Ham_{\anh}^3\rangle_{\Ham_\harm}-3\langle \Ham_{\anh}^2\rangle_{\Ham_\harm}\langle \Ham_{\anh}\rangle_{\Ham_\harm}+2\langle \Ham_{\anh}\rangle^3_{\Ham_\harm}\nonumber.
\end{align*}}. Obviously, since the splitting of the anharmonic and harmonic parts is essentially arbitrary, the \emph{complete} infinite series of the above expansion should not be a function of the unknown harmonic potential $\mathcal{V}_\har$.  However, due to finite truncation, in practice, the sum of the series has a dependence on this potential. This then leads to the variational principle of minimal sensitivity which allows us to optimize the approximation i.e., 
\begin{eqnarray} \label{eq:min-sens}
\begin{aligned}
 \frac{\partial \mathcal{F}}{{ \partial }\mathcal{V}_\har} =0 \quad \longrightarrow \quad 
    \frac{\partial \mathcal{F}}{\partial M^2} {=}0.
\end{aligned}
\end{eqnarray}

The free energy of the harmonic part of the effective Hamiltonian $\Ham_\harm$, is calculated by substituting \eqref{eq-20} into \eqref{eq:free-energy} as below
{\small
\begin{equation} \label{eq:f_tr_experssion}
\begin{aligned}
\mathcal{F}_{\harm}&=
\mathcal{O} + \sum _{\bfk\in \mathbb{K}} \frac{1}{2\beta }\log \left[\frac{L^2 {\left| {\bfk}\right| }^4 \left( {\left| {\bfk}\right| } \kappa_b -2 e^{\left| \bfk\right| d} \epsilon_{0}  \left(f^2-a_z \kappa_b \right)\right)}{{\left| {\bfk}\right| }+2 a_z e^{\left| \bfk\right| d} \epsilon_{0} }+ \frac{L^2}{2}m^4 
\right]\\
&=\mathcal{O} +\frac{L^2}{4\pi ^2}\int_{2\pi/L}^{2\pi/a}  \frac{1}{2\beta }\log\left[\frac{L^2 {k }^4 \left({ {k} } \kappa_b -2 e^{k d} \epsilon_{0}  \left(f^2-a_z \kappa_b \right)\right)}{{k }+2 a_z e^{\left| \bfk\right| d} \epsilon_{0} }+
 \frac{L^2}{2}m^4 
\right]2\pi k \mathrm{d}{k }\\
&\simeq \mathcal{O} +\frac{L^2}{4\pi ^2}\int_{2\pi/L}^{2\pi/a}  \frac{1}{2\beta }\log \left[
\kappa _{\eff} k^4 +
\frac{1}{2}m^4
\right]2\pi  k \mathrm{d} k=\frac{L^2 }{8\beta } \frac{m^2 }{\sqrt{2\kappa _{\eff}}} ,
\end{aligned}
\end{equation}}
where $\mathcal{O}$ is constant of no consequence and $\kappa _{\eff}= \kappa_b -\frac{f^2}{a_z}$ illustrating a result already known~\cite{deng2014flexoelectricity, mohammadi2014theory} that flexoelectricity causes a renormalization of the bending modulus. By substituting the anharmonic Hamiltonian in \eqref{eq:h_int} in the free energy of the system in \eqref{eq:Finf} we obtain~\cite{bachmann1999strong}
\begin{equation} \label{eq:F_expansion}
\begin{aligned}
\mathcal{F}&=\mathcal{F}_\po-\frac{1}{\beta }\log [\mathcal{Z}_v]=\mathcal{F}_\po+\mathcal{F}_\har
-\frac{1}{\beta}\sum_{n=1}^\infty\frac{(-g\beta/2)^n}{n!} \bigg\langle \left( \int_\SU {V_{\anh}} \right)^n\bigg\rangle^c_{\Ham_\harm}\\
%
&=\mathcal{F}_\po+\frac{L^2 }{8\beta } m^2\frac{1}{\sqrt{2\kappa _{\eff}}} +m^2 \sum_{n=1}^{\infty}\gamma_n \left( \frac{g}{m^2} \right)^n,
\end{aligned}
\end{equation}
where coefficients $\gamma_i$ can be obtained from expanding the free energy in \eqref{eq:Finf} with respect to $g$ as following
\begin{equation} \label{eq:a_i}
\begin{aligned}
\gamma_1&= \frac{m^4}{2} \alpha _4\int_\SU \langle v^4(\bfx)\rangle^c_{\Ham_\harm}, \\
\\
\gamma_2&=\frac{m^6}{2}\alpha _6\int_\SU \langle v^6(\bfx)\rangle^c_{\Ham_\harm} - \frac{m^{10}}{8}{\beta }\alpha _4^2 \int_\SU \langle v^4(\bfx)v(\bfx')^4\rangle^c_{\Ham_\harm}.
\end{aligned}
\end{equation}

We remark that an inspection of Equations \eqref{eq:F_expansion} and \eqref{eq:a_i} shows the necessity to calculate  ensemble averages to obtain the free energy of the system. By Wick's theorem, we can expand the higher order ensemble averages as follows
\begin{equation} \label{eq:wick1}
\begin{aligned}
&\langle v^4(\bfx)\rangle_{\Ham_\harm}^c  = 3\langle v^2(\bfx)\rangle_{\Ham_\harm}^2,\\
&\langle v^6(\bfx)\rangle_{\Ham_\harm}^c   =15\langle  v^2(\bfx)\rangle_{\Ham_\harm} ^3.
 \end{aligned}
\end{equation} 

The ensemble averages of the square of the height function in real and Fourier space can be derived to be
\begin{eqnarray} \label{eq-22}
\begin{aligned}
\langle  |\hat{v}(\bfk)|^2 \rangle_{\Ham_\harm} &=\frac{\int  |\hat{v}(\bfk)|^2e^{-\beta \Ham_\harm}\prod _{\bfk \in \mathbb{K}}}{\int
e^{-\beta \Ham_\harm} \, \prod _{\bfk \in \mathbb{K}}}=
\frac{1}{ L^2 \beta  \left(m^4+2\kappa _{\eff}|\bfk|^4 \right)},\\
\\
\langle v^2\rangle_{\Ham_\harm}  &=\frac{1}{L^2}\int_\SU \langle v^2(\bfx)\rangle_{\Ham_\harm} =\sum _{\bfk\in \mathbb{K}} \langle  |\hat{v}(\bfk)|^2\rangle_{\Ham_\harm} =\sum _{\bfk\in \mathbb{K}} \frac{1}{ L^2 \beta \left(m^4+2\kappa _{\eff}|\bfk|^4 \right)}\\
&\simeq \frac{L^2}{4\pi ^2}\int_{2\pi /L}^{2\pi /a} \frac{2\pi  k }{ L^2 \beta  \left(m^4+2\kappa _{\eff}k^4 \right)} \, \mathrm{d}k 
=\frac{1}{8\beta m^2 \sqrt{2\kappa _{\eff}}}.
\end{aligned}
\end{eqnarray}

Also, by the use of Wick's theorem, the higher order correlations can be expressed in terms of the lower order correlations which can be obtained by the use of Feynman diagrams~\cite{bachmann1999strong}
{\small
\begin{equation} \label{eq:wick2}
\begin{aligned}
\int _{\SU}\langle v^4(\bfx)v(\bfx')^4\rangle_{\Ham_\harm}^c &=72\int _{\SU}\langle \left| v(\bfx)\right| ^2\rangle_{\Ham_\harm} \langle \left| v(\bfx')\right| ^2\rangle_{\Ham_\harm} \langle v(\bfx)v(\bfx')\rangle_{\Ham_\harm} ^2
+24\int _{\SU}\langle v(\bfx)v(\bfx')\rangle_{\Ham_\harm} ^4\\
 &=\frac{21L^2}{512 \sqrt{2} \beta ^4 m^{10} \kappa _{\text{eff}}^{3/2}}.
 \end{aligned}
\end{equation}
}\\

The free energy ${\mathcal{F}} $ can be obtained by substituting Equations \eqref{eq:wick1}-\eqref{eq:wick2} into Equation~\eqref{eq:Finf}. Our ultimate goal is to impose the hard wall constraint by assuming $m\to 0$. This limit must be handled carefully and we adopt the following procedure. Assuming $r$ is an independent variable of $g$ and expanding $m^2=\sqrt{M^4-g r}$ in powers of $g$, in the limit of $m\to 0$, we obtain~\cite{duttmann2009variational}
\begin{equation} \label{eq-25}
\begin{aligned}
m^2=\sqrt{M^4-g r}=M^2-\frac{1}{2}\frac{r}{M^2}g-\frac{1}{8}\frac{r^2}{M^6}g^2+O(g^3).
 \end{aligned}
\end{equation}

Substituting this expansion back in our free energy in Equation~\eqref{eq:F_expansion}, assuming $r\rightarrow \frac{M^4-m^4}{g}$
and $m \rightarrow 0$ (hard wall constraint), the free energy is obtained as~\cite{bachmann1999strong}
\begin{equation} \label{eq-26}
\begin{aligned}
\mathcal{F}_N&=\mathcal{F}_\po+\mathcal{F}_\har b_0^{(N)}+\sum _{n=1}^N a_ng^nM^{2(1-n)}b_n^{(N)}\\
&=\mathcal{F}_\po+M^2\frac{L^2}{8\beta \sqrt{\kappa _{\eff}}}b_0^N+\sum _{n=1} \gamma_n b_n^{(N)} g^nM^{2(1-n)},
 \end{aligned}
\end{equation} 
where $b_n^{(N)}$ is defined as\:
\begin{equation} \label{eq-27}
\begin{aligned}
 b_n^{(N)}&=\sum _{k=0}^{N-n} (-1)^k\left(
\begin{array}{c}
 \frac{1-n}{2} \\
 k \\
\end{array}
\right).
\end{aligned}
\end{equation}

\subsection{Approximation of the entropic pressure assuming a penalty potential up to quartic order}

Up to a quartic order in the penalty potential, the free energy can be approximated as
\begin{equation} \label{eq-29}
\begin{aligned}
\mathcal{F}_1&=\mathcal{F}_\po+\mathcal{F}_\har+\frac{b_1^{(1)}}{2}\int _{\SU}g\left(M^2\right)^2\alpha _4\langle v^4(\bfx)\rangle^c_{\Ham_\harm}\\
\\
&=\mathcal{F}_\po+M^2\frac{L^2}{8\beta }\frac{b_0^{(1)}}{\sqrt{2\kappa _{\eff}}} +\frac{3}{2}L^2 b_1^{(1)} gM^4\alpha _4\left(\frac{1}{8M^2\beta \sqrt{2\kappa _{\eff}}}\right)^2\\
\\
&=\mathcal{F}_\po+M^2\frac{L^2}{8\beta }\frac{b_0^{(1)}}{\sqrt{2\kappa _{\eff}}}+\frac{g L^2 b_1^{(1)}}{128 \beta ^2 \kappa _{\eff}}.
\end{aligned}
\end{equation}

Since the free energy should be independent of M, the truncation should have the minimum sensitivity. Therefore,
\begin{equation} \label{eq-30}
\begin{aligned}
\frac{\partial \mathcal{F}_1}{\partial M^2}{=}0 \quad \longrightarrow \quad   \frac{L^2 }{8\beta } \frac{b_0^{(1)}}{\sqrt{2\kappa _{\eff}}}{=}0.
\end{aligned}
\end{equation}

There is no solution for $M^2$, therefore we assume $ M^2\to 0$. By this assumption, the free energy becomes
\begin{equation} \label{eq-31}
\begin{aligned}
\mathcal{F}_1=\mathcal{F}_\po+\frac{g L^2  b_1^{(1)}}{128  \beta ^2 \kappa _{\eff}}=\mathcal{F}_\po+\frac{\pi ^2 L^2 }{128 d^2\beta ^2 \kappa _{\eff}}.
\end{aligned}
\end{equation}

The free energy contribution from the polarization($\mathcal{F}_\po$) can be obtained from Equation~\eqref{eq:c_a}
\begin{eqnarray} \label{eq:F_p}
\begin{aligned}
\mathcal{F}_\po&=-\frac{1}{\beta}\log(\mathcal{Z}_\po)=\sum_{\bfk \in \mathbb{K}} \frac{1}{2\beta }\log \left[4a_z^2-\frac{e^{-2\left| \bfk\right| d}\left| \bfk\right|^2 }{\epsilon_{0}^2}\right]\\
&=\frac{L^2}{4\pi ^2}\int \frac{1}{2\beta }\log \left[4a_z^2-\frac{e^{-2 k d}k^2 }{\epsilon_{0} ^2}\right]2\pi k \mathrm{d}k\\
&=\mathcal{O} +\frac{L^2}{4\pi \beta } \int_{2\pi/L}^{2\pi /a}\log \left[4a_z^2\epsilon_{0}^2-e^{-2k d}k^2 \right] k \mathrm{d}k,
\end{aligned}
\end{eqnarray} 
where $\mathcal{O} $ is constant of no consequence. Following \eqref{eq:pressure}, we find the entropic pressure to be
\begin{equation} \label{eq-33}
\begin{aligned}
\mathcal{P}&=-\frac{1}{4L^2}\left(\frac{\partial  \mathcal{F}_1}{\partial d}\right)=\frac{  \pi ^2 }{256 d^3  \beta ^2 \kappa _{\eff}}
-
\frac{\partial }{\partial d}\left(\frac{1}{16\beta  \pi }\int_{2\pi/L}^{2\pi/a} \log \left[4a_z^2\epsilon_{0} ^2-e^{-2kd}k^2 \right] k \mathrm{d}k\right)\\
&=\frac{  \pi ^2 }{256 d^3 \beta ^2 \kappa _{\eff}}
-
\frac{1}{16\beta  \pi }\int_{2\pi/L}^{2\pi/a} \frac{\partial }{\partial d}\left(\log \left[4a_z^2\epsilon_{0}^2-e^{-2k d}k^2 \right] k\right) \mathrm{d}k\\
&\simeq \frac{  \pi ^2 }{256 d^3  \beta ^2 \kappa _{\eff}}
-
\frac{1}{8\beta  \pi }\int_{2\pi/L}^{2\pi/a} \frac{k^4}{4e^{2 d k} a_z^2 \epsilon_{0}^2} \mathrm{d}k  \\
%
&\simeq\begin{cases}
 \frac{  \pi ^2 }{256 d^3  \beta ^2 \kappa _{\eff}}-\frac{3}{128 a_z^2 d^5 \pi  \beta \epsilon_{0} ^2}  e^{-4 \pi  d/L} \qquad \text{large distances} \\
\frac{  \pi ^2 }{256 d^3  \beta ^2 \kappa _{\eff}}-\frac{3}{128 a_z^2 d^5 \pi  \beta  \epsilon_{0} ^2} \qquad \qquad \quad  \text{short distances}
\end{cases}
\end{aligned}
\end{equation} \\

\subsection{Approximation of the entropic pressure assuming a penalty potential up to a hexic order}

We may go a step further to ensure higher accuracy and approximate the free energy up to the hexic order as follows
\begin{equation} \label{eq-34}
\begin{aligned}
\mathcal{F}_2=&\mathcal{F}_\po+
M^2\frac{L^2}{8\beta }\frac{b_0^{(2)}}{\sqrt{2\kappa _{\eff}}}+
\frac{b_1^{(2)}}{2}\int _{\SU}gM^4\alpha_4\langle v^4(\bfx)\rangle^c_{\Ham_\harm}  +\frac{b_2^{(2)}}{2}\int _{\SU}g^2M^4\alpha _6\langle v^6(\bfx)\rangle^c_{\Ham_\harm}
\\
& - \frac{\beta }{8} b_2^{(2)}\int _{\SU}g^2M^8\alpha _4^2  \langle v^4(\bfx)v(\bfx')^4\rangle^c_{\Ham_\harm} \\
=&\mathcal{F}_\po+
M^2\frac{L^2}{8\beta }\frac{b_0^{(2)}}{\sqrt{2\kappa _{\eff}}}+
b_1^{(2)} g\left(\frac{1}{8\beta  \sqrt{2\kappa _{\eff}}}\right)^2+\frac{17}{6}b_2^{(2)}g^2\frac{1}{M^2} \left(\frac{1}{8\beta  \sqrt{2\kappa _{\eff}}}\right)^3 
\\
&-\frac{1}{18} \beta  b_2^{(2)}g^2L^2\frac{21}{512 \sqrt{2} M^{2} \beta ^4 \kappa _{\eff}^{3/2}}.
\end{aligned}
\end{equation}  
\\
Minimizing the sensitivity of this truncated free energy to the variational parameter $M$ we obtain $M$ as
\begin{equation} \label{eq-35}
\begin{aligned}
\frac{\partial \mathcal{F}_2}{\partial M^2}{=}0
\quad \longrightarrow \quad   M^2= 
 \frac{g}{16 \beta }
\sqrt{\frac{b_2^{(2)}}{b_0^{(2)} \kappa _\eff}} .
\end{aligned}
\end{equation}  

Hence, the free energy yields as
\begin{equation} \label{eq-36}
\begin{aligned}
\mathcal{F}_2=\mathcal{F}_\po+ \frac{\left(\sqrt{3}+2\right) \pi ^2}{256 \beta ^2 d^2 \kappa _{\text{eff}}}.
\end{aligned}
\end{equation}  

The entropic pressure is then obtained as
\begin{equation} \label{eq-38}
\begin{aligned}
\mathcal{P}&=-\frac{1}{4L^2}\left(\frac{\partial \mathcal{F}_2}{\partial d}\right)\\
&=
{ \frac{\left(\sqrt{3}+2\right) \pi ^2}{512 \beta ^2 d^3 \kappa_\eff}}
- 
\frac{\partial }{\partial d}\left(\frac{1}{16\beta  \pi }\int_{2\pi/L}^{2\pi/a} \log \left[4a_z^2\epsilon_{0} ^2-e^{-2kd}k^2 \right] k \mathrm{d}k\right)\\
&
\simeq 
{ \frac{\left(\sqrt{3}+2\right) \pi ^2}{512 \beta ^2 d^3 \kappa_\eff}}
-
\frac{1}{8\beta  \pi }\int_{2\pi/L}^{2\pi/a} \frac{k^4}{4e^{2 d k} a_z^2 \epsilon_{0}^2} \mathrm{d}k  \\
&\simeq\begin{cases}
{ \frac{\left(\sqrt{3}+2\right) \pi ^2}{512 \beta ^2 d^3 \kappa_\eff}}
-
\frac{3}{128 a_z^2 d^5 \pi  \beta  \epsilon_{0} ^2}  e^{-4 \pi  d/L} \qquad \text{large distances} \\
{ \frac{\left(\sqrt{3}+2\right) \pi ^2}{512 \beta ^2 d^3 \kappa_\eff}}
-
\frac{3}{128 a_z^2 d^5 \pi  \beta  \epsilon_{0} ^2}  \qquad \qquad \quad  \text{short distances}
\end{cases}
\end{aligned}
\end{equation}

\section{Results and Discussion}\label{sec:applications}

\subsection{Accuracy of our model and comparison with other works}

Before proceeding to present our results, we comment on our choice of model parameters. The typical bending modulus of the biological membranes varies in the range of $\kappa=15-25 k_BT$~\cite{raphael2000membrane}. Also, the typical thickness of the biological membranes is reported as $t=4-8$ nm~\cite{bivas1981flexoelectric}. The well-known flexoelectric constant $f_e:=-f/a_z$ of the biological membranes is predicted to take value of $f_e=0.3-150 \times 10^{-19} $ C~\cite{petrov1998mechanosensitivity,petrov2002flexoelectricity,liu2013flexoelectricity}. We assume that the temperature in the biological temperature of the human body($T=310^\circ $K) and $\epsilon=2\epsilon_0$~\cite{ahmadpoor2015thermal}. \\

As shown earlier by~\cite{bachmann1999strong}, the series expansion with the variational perturbation procedure in the preceding section leads to a convergent series as opposed to a naive perturbation method. The assessment of the accuracy of our results can simply be done by going one step further in the free energy expansion. Also, comparison between the entropic pressure obtained by our method and the entropic pressure obtained by Monte Carlo simulations by Janke and Kleinert~\cite{janke1986fluctuation} confirms the accuracy of our model. \\

Figure~\ref{fig:comparison} shows a comparison of the entropic pressure obtained from various models including Helfrich's original work that excluded flexoelectricity~\cite{helfrich1984undulations,helfrich1978steric}, Bachmann's model~\cite{bachmann1999strong, kleinert1998pressure} which (like ours) is also based on a variational perturbation approach but excluded flexoelectricity, and our model which accounts for both flexoelectricity and polarization. We do not present the model of Bivas and Petrov~\cite{bivas1981flexoelectric} in the figure since they only presented asymptotic limits. We simply remark that qualitatively, our model confirms their asymptotic limit. Upon ignoring electrostatic effects (flexoelectricity and polarization), our constructed model identically matches Bachmann's model. Note that for inter-membrane distances smaller that $0.5$ nm, our final estimation of Equation \eqref{eq-38} is supposed to become less precise. Accordingly, we performed numerical integration to obtain the exact behavior of the entropic pressure for $d<0.5$ nm and found that our approximation of Equation \eqref{eq-38} matches well with numerical integration of the pressure for $d>0.5$ nm. Our main conclusion is that flexoelectricity induces a strong attraction for small inter-membrane separations and increases the repulsion for very larger separations. Thus, flexoelectricity acts as an ``amplifier". Interestingly, the role of flexoelectricity as an amplifier is also observed in the context of the mechanics of hair cells and the hearing mechanism~\cite{breneman2009hair, deng2019collusion}. The amplification behavior may be rationalized by examining the expression derived earlier for the force. Electrostatics appears to impact the entropic pressure in two ways: (1) The bending modulus is softened i.e., $\kappa _{\eff}= \kappa_b -\frac{f^2}{a_z}$ and hence the entropic repulsion is amplified. There is a second term that is attractive in nature but varies as $1/d^5$ for short distances and decays exponentially at larger separations.

\begin{figure}[hbt!]
    \centering
    \includegraphics[width=0.65
\linewidth]{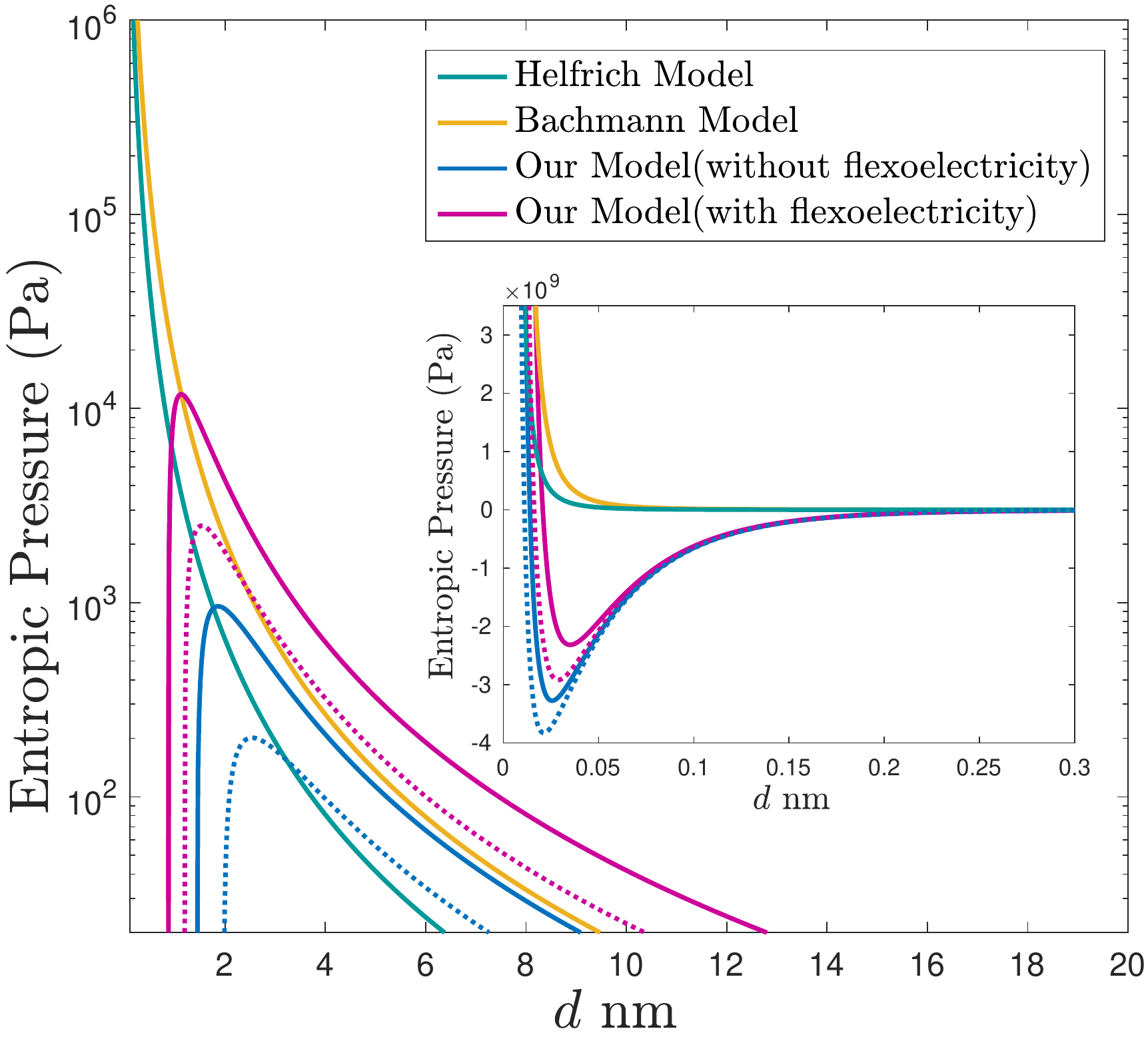}
    \caption{\small Comparison of Helfrich's model~\cite{helfrich1984undulations,helfrich1978steric} and Bachmann's model~\cite{bachmann1999strong, kleinert1998pressure} with our work(up to hexic order). The blue line indicates our model without accounting for flexoelectricity which matches Bachmann's model at large distances where the polarization effect fades. { The dotted lines are the same approximations where the penalty potential is up to quartic order from Equation~\eqref{eq-33}.} The numerical parameters for this plot are $\kappa=25 k_B T$, $t=4$ nm, $\epsilon=2\epsilon_0$ and $f=-0.31 \times 10^{-19} a_z$ NM/C. }
    \label{fig:comparison}
\end{figure}

\subsection{Comparison with van der Waals and hydration forces}

As we highlighted in the introduction, in addition to the entropic repulsion (due to purely mechanical origins), the other two main forces (for electrically neutral membranes) are the hydration and van der Waals. Therefore, the net interacting potential per unit area can be written as~\cite{ lipowsky1986unbinding}
\begin{equation}
    V_{net}(d)=V_H(d)+V_W(d)+V_s(d),
\end{equation}
where, $V_H(d)$, $V_W(d)$ and $V_s(d)$ stand for hydration, van der Waals and entropic interaction potentials. The hydration potential can be written as~\cite{ lipowsky1986unbinding}
\begin{equation}
    V_H(d)=A_H \exp[-d/\lambda_H],
\end{equation}
where $A_H \simeq 0.2$ J$/$m$^2$ and $\lambda_H  \simeq 0.3$ nm denote the hydration strength and length-scale. A good approximation for the van der Waals potential between two interacting membranes is~\cite{ lipowsky1986unbinding}
\begin{equation}
    V_W(d)=-\frac{W_H}{12 \pi} \left( \frac{1}{d^2}-\frac{2}{(d+t)^2}+\frac{1}{(d+2t)^2} \right),
\end{equation}
where $W_H \simeq 0.61-8.2 \times 10^{-21}$ J is the Hamaker constant. The hydration repulsion is very short ranged and the cut off length is around few angstroms~\cite{helfrich1984undulations,mutz1989unbinding}. For specific strength of entropic repulsion and van der Waals interaction, we can expect to have up to two stable states which indicate bounded and totally separated states~\cite{helfrich1984undulations}.\\

\begin{figure}[hbt!]
    \centering
    \includegraphics[width=0.65
\linewidth]{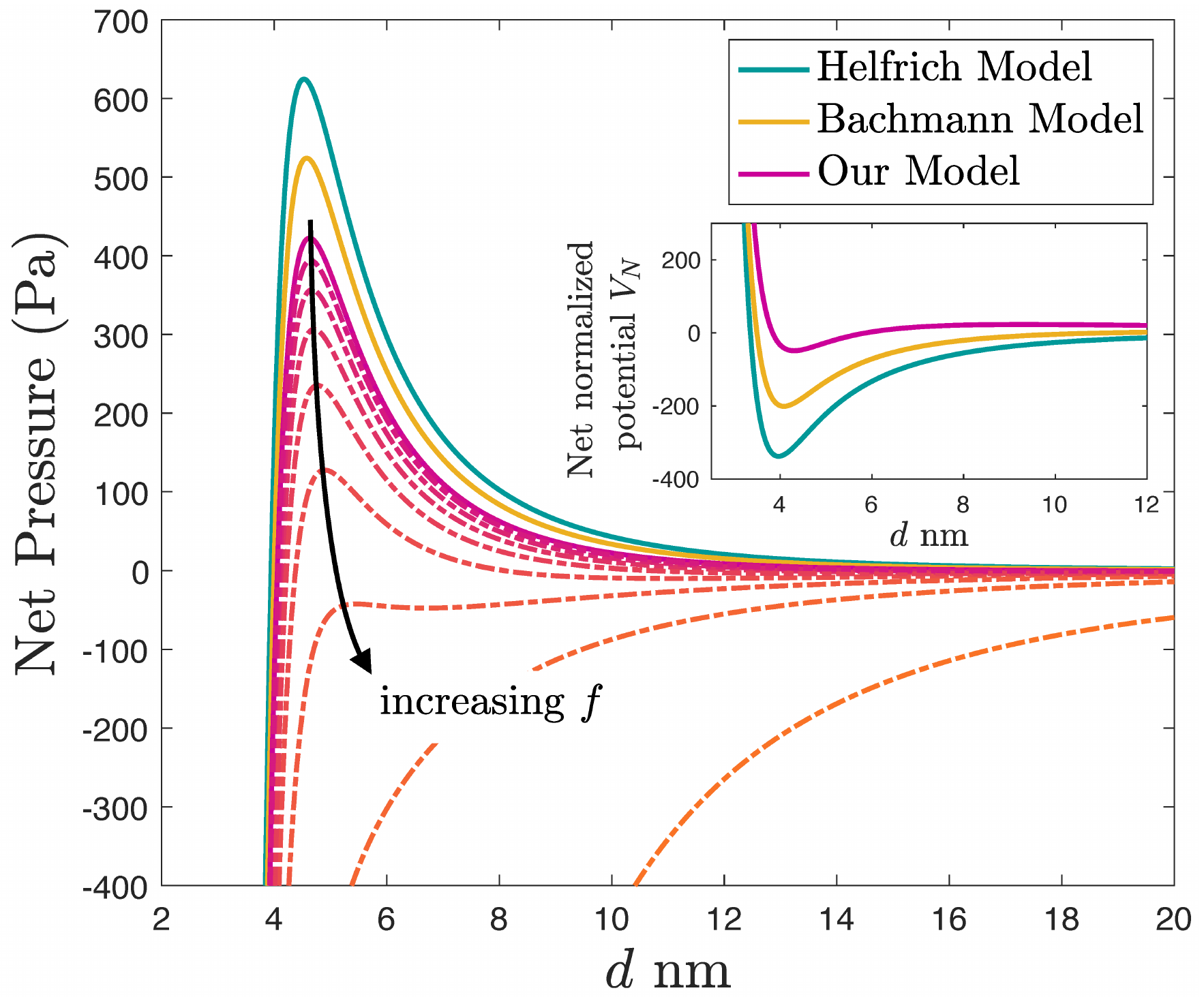}
    \caption{\small The net pressure on each of the membranes in proximity of another membrane that are apart with distance of $d$. The numerical values are $\kappa=25 k_B T$, $t=4$ nm, $\epsilon=2\epsilon_0$, $f= - 0.31 \times 10^{-19} a_z$ NM/C and $W_H=2 \times 10^{-21}$ J. The inset is the net normalized potential $V_N=V_\tot \times L^2/k_BT$ with respect to the inter-membrane distance $d$ to highlight the existence of a stable state around inter-membrane distance of $4$ nm. }
    \label{fig:netforce}
\end{figure}

For obtaining an idea about the binding and unbinding transition between two membranes, we add up the relevant forces and obtain the equilibrium state\footnote{It is worthwhile to mention that it is not appropriate to just add up the forces due to different origins. A systematic statistical mechanics study must be done ab initio. However, for qualitative conclusions, such an approximation is certainly worthwhile. More accurate estimation of the equilibrium state have been done through renormalization based approaches~\cite{lu2015effective,mutz1989unbinding,helfrich1984undulations}.}. In Figure~\ref{fig:netforce}, we illustrate the net force (per area) on each of the membranes in a pair membrane system. As in the previous sub-section, our main conclusion is that flexoelectricity and polarization have a meaningful impact primarily at small inter-membrane separations or when quite further apart. In those regimes, the electrostatic contribution can indeed become competitive with other forces. From the viewpoint of unbinding-binding transition and adhesion, we also show the potential energy in the inset to highlight how the energy minima is altered due to flexoelectricity as well as its spatial location.\\

\section{Concluding remarks}\label{sec:Conclusions}
\label{sec-conclusion}

In conclusion, we have obtained an approximate but highly accurate closed-form solution to the entropic force between two fluctuating flexoelectric membranes. We find an enhanced attractive force for close-membrane separations and an enhanced repulsion for larger separations. A detailed physical consequence of our calculation on biology is beyond the scope of the present work but we anticipate its utility in nearly all the phenomena described in the topical paragraph of this paper. It remains to be seen what consequence our work will have (or not) on those problems.\\

\bibliographystyle{elsarticle-num}
\bibliography{References.bib} 

\begin{thebibliography}{10}
\expandafter\ifx\csname url\endcsname\relax
  \def\url#1{\texttt{#1}}\fi
\expandafter\ifx\csname urlprefix\endcsname\relax\def\urlprefix{URL }\fi
\expandafter\ifx\csname href\endcsname\relax
  \def\href#1#2{#2} \def\path#1{#1}\fi

\bibitem{pastan1985pathway}
I.~Pastan, M.~C. Willingham, The pathway of endocytosis, in: Endocytosis,
  Springer, 1985, pp. 1--44.

\bibitem{fisher1993force}
L.~Fisher, Force between biological surfaces, Journal of the Chemical Society,
  Faraday Transactions 89~(15) (1993) 2567--2582.

\bibitem{chernomordik1995lipids}
L.~Chernomordik, M.~M. Kozlov, J.~Zimmerberg, Lipids in biological membrane
  fusion, The Journal of membrane biology 146~(1) (1995) 1--14.

\bibitem{lipowsky1991adhesion}
R.~Lipowsky, U.~Seifert, Adhesion of vesicles and membranes, Molecular crystals
  and liquid crystals 202~(1) (1991) 17--25.

\bibitem{lipowsky1987erratum}
R.~Lipowsky, S.~Leibler, Erratum: Unbinding transitions of interacting
  membranes [phys. rev. lett. 56, 2541 (1986)], PhRvL 59~(17) (1987) 1983.

\bibitem{helfrich1978steric}
W.~Helfrich, Steric interaction of fluid membranes in multilayer systems,
  Zeitschrift f{\"u}r Naturforschung A 33~(3) (1978) 305--315.

\bibitem{bonazzi2011impenetrable}
M.~Bonazzi, P.~Cossart, Impenetrable barriers or entry portals? the role of
  cell--cell adhesion during infection, Journal of Cell Biology 195~(3) (2011)
  349--358.

\bibitem{helfrich1984undulations}
W.~Helfrich, R.-M. Servuss, Undulations, steric interaction and cohesion of
  fluid membranes, Il Nuovo Cimento D 3~(1) (1984) 137--151.

\bibitem{ninham1970van}
B.~Ninham, V.~Parsegian, Van der waals interactions in multilayer systems, The
  Journal of Chemical Physics 53~(9) (1970) 3398--3402.

\bibitem{rand1989hydration}
R.~Rand, V.~Parsegian, Hydration forces between phospholipid bilayers,
  Biochimica et Biophysica Acta (BBA)-Reviews on Biomembranes 988~(3) (1989)
  351--376.

\bibitem{lipowsky1995structure}
R.~Lipowsky, E.~Sackmann, Structure and dynamics of membranes: I. from cells to
  vesicles/II. generic and specific interactions, Elsevier, 1995.

\bibitem{israelachvili1992entropic}
J.~N. Israelachvili, H.~Wennerstroem, Entropic forces between amphiphilic
  surfaces in liquids, The Journal of Physical Chemistry 96~(2) (1992)
  520--531.

\bibitem{milner1992flory}
S.~Milner, D.~Roux, Flory theory of the unbinding transition, Journal de
  Physique I 2~(9) (1992) 1741--1754.

\bibitem{lipowsky1986unbinding}
R.~Lipowsky, S.~Leibler, Unbinding transitions of interacting membranes,
  Physical Review Letters 56~(23) (1986) 2541.

\bibitem{lipowsky1989binding}
R.~Lipowsky, B.~Zielinska, Binding and unbinding of lipid membranes: A monte
  carlo study, Physical Review Letters 62~(13) (1989) 1572.

\bibitem{hanlumyuang2014revisiting}
Y.~Hanlumyuang, L.~Liu, P.~Sharma, Revisiting the entropic force between
  fluctuating biological membranes, Journal of the Mechanics and Physics of
  Solids 63 (2014) 179--186.

\bibitem{bachmann2001fluctuation}
M.~Bachmann, H.~Kleinert, A.~Pelster, Fluctuation pressure of a stack of
  membranes, Physical Review E 63~(5) (2001) 051709.

\bibitem{lu2015effective}
B.-S. Lu, R.~Podgornik, Effective interactions between fluid membranes,
  Physical Review E 92~(2) (2015) 022112.

\bibitem{janke1986fluctuation}
W.~Janke, H.~Kleinert, Fluctuation pressure of membrane between walls, Physics
  Letters A 117~(7) (1986) 353--357.

\bibitem{sharma2013entropic}
P.~Sharma, Entropic force between membranes reexamined, Proceedings of the
  National Academy of Sciences 110~(6) (2013) 1976--1977.

\bibitem{freund2013entropic}
L.~Freund, Entropic pressure between biomembranes in a periodic stack due to
  thermal fluctuations, Proceedings of the National Academy of Sciences 110~(6)
  (2013) 2047--2051.

\bibitem{liang2016fluctuating}
X.~Liang, P.~K. Purohit, A fluctuating elastic plate and a cell model for lipid
  membranes, Journal of the Mechanics and Physics of Solids 90 (2016) 29--44.

\bibitem{schneider1984thermal}
M.~Schneider, J.~Jenkins, W.~Webb, Thermal fluctuations of large
  quasi-spherical bimolecular phospholipid vesicles, Journal de Physique 45~(9)
  (1984) 1457--1472.

\bibitem{morse1994fluctuations}
D.~Morse, S.~T. Milner, Fluctuations and phase behavior of fluid membrane
  vesicles, EPL (Europhysics Letters) 26~(8) (1994) 565.

\bibitem{morse1995statistical}
D.~C. Morse, S.~T. Milner, Statistical mechanics of closed fluid membranes,
  Physical Review E 52~(6) (1995) 5918.

\bibitem{michalet1994fluctuating}
X.~Michalet, D.~Bensimon, B.~Fourcade, Fluctuating vesicles of nonspherical
  topology, Physical review letters 72~(1) (1994) 168.

\bibitem{seifert1995concept}
U.~Seifert, The concept of effective tension for fluctuating vesicles,
  Zeitschrift f{\"u}r Physik B Condensed Matter 97~(2) (1995) 299--309.

\bibitem{bachmann1999strong}
M.~Bachmann, H.~Kleinert, A.~Pelster, Strong-coupling calculation of
  fluctuation pressure of a membrane between walls, Physics Letters A 261~(3-4)
  (1999) 127--133.

\bibitem{gompper1989steric}
G.~Gompper, D.~Kroll, Steric interactions in multimembrane systems: a monte
  carlo study, EPL (Europhysics Letters) 9~(1) (1989) 59.

\bibitem{janke1987fluctuation}
W.~Janke, H.~Kleinert, Fluctuation pressure of a stack of membranes, Physical
  review letters 58~(2) (1987) 144.

\bibitem{liang2018method}
X.~Liang, P.~K. Purohit, A method to compute elastic and entropic interactions
  of membrane inclusions, Extreme mechanics letters 18 (2018) 29--35.

\bibitem{chen2015entropic}
D.~Chen, Y.~Kulkarni, Entropic interaction between fluctuating twin boundaries,
  Journal of the Mechanics and Physics of Solids 84 (2015) 59--71.

\bibitem{chen2017thermal}
D.~Chen, Y.~Kulkarni, Thermal fluctuations as a computational microscope for
  studying crystalline interfaces: A mechanistic perspective, Journal of
  Applied Mechanics 84~(12) (2017).

\bibitem{krichen2016flexoelectricity}
S.~Krichen, P.~Sharma, Flexoelectricity: A perspective on an unusual
  electromechanical coupling, Journal of Applied Mechanics 83~(3) (2016).

\bibitem{tagantsev1986piezoelectricity}
A.~Tagantsev, Piezoelectricity and flexoelectricity in crystalline dielectrics,
  Physical Review B 34~(8) (1986) 5883.

\bibitem{zubko2013flexoelectric}
P.~Zubko, G.~Catalan, A.~K. Tagantsev, Flexoelectric effect in solids, Annual
  Review of Materials Research 43 (2013).

\bibitem{nguyen2013nanoscale}
T.~D. Nguyen, S.~Mao, Y.-W. Yeh, P.~K. Purohit, M.~C. McAlpine, Nanoscale
  flexoelectricity, Advanced Materials 25~(7) (2013) 946--974.

\bibitem{ahmadpoor2015flexoelectricity}
F.~Ahmadpoor, P.~Sharma, Flexoelectricity in two-dimensional crystalline and
  biological membranes, Nanoscale 7~(40) (2015) 16555--16570.

\bibitem{mao2014insights}
S.~Mao, P.~K. Purohit, Insights into flexoelectric solids from strain-gradient
  elasticity, Journal of Applied Mechanics 81~(8) (2014).

\bibitem{petrov1993flexoelectric}
A.~G. Petrov, B.~A. Miller, K.~Hristova, P.~N. Usherwood, Flexoelectric effects
  in model and native membranes containing ion channels, European biophysics
  journal 22~(4) (1993) 289--300.

\bibitem{petrov1996flexoelectricity}
A.~Petrov, M.~Spassova, J.~Fendler, Flexoelectricity and photoflexoelectricity
  in model and biomembranes, Thin Solid Films 284 (1996) 845--848.

\bibitem{petrov1998mechanosensitivity}
A.~G. Petrov, Mechanosensitivity of cell membranes: role of liquid crystalline
  lipid matrix, in: Liquid Crystals: Chemistry and Structure, Vol. 3319,
  International Society for Optics and Photonics, 1998, pp. 306--318.

\bibitem{petrov2002flexoelectricity}
A.~G. Petrov, Flexoelectricity of model and living membranes, Biochimica et
  Biophysica Acta (BBA)-Biomembranes 1561~(1) (2002) 1--25.

\bibitem{petrov2006electricity}
A.~G. Petrov, Electricity and mechanics of biomembrane systems:
  flexoelectricity in living membranes, Analytica chimica acta 568~(1-2) (2006)
  70--83.

\bibitem{petrov2007flexoelectricity}
A.~G. Petrov, Flexoelectricity and mechanotransduction, Current Topics in
  Membranes 58 (2007) 121--150.

\bibitem{raphael2000membrane}
R.~M. Raphael, A.~S. Popel, W.~E. Brownell, A membrane bending model of outer
  hair cell electromotility, Biophysical journal 78~(6) (2000) 2844--2862.

\bibitem{spector2006electromechanical}
A.~Spector, N.~Deo, K.~Grosh, J.~Ratnanather, R.~Raphael, Electromechanical
  models of the outer hair cell composite membrane, The Journal of membrane
  biology 209~(2-3) (2006) 135--152.

\bibitem{breneman2009piezo}
K.~D. Breneman, R.~D. Rabbitt, Piezo-and flexoelectric membrane materials
  underlie fast biological motors in the ear, in: Materials Research Society
  symposia proceedings. Materials Research Society, Vol. 1186, NIH Public
  Access, 2009.

\bibitem{brownell2010cell}
W.~E. Brownell, F.~Qian, B.~Anvari, Cell membrane tethers generate mechanical
  force in response to electrical stimulation, Biophysical journal 99~(3)
  (2010) 845--852.

\bibitem{glassinger2005electromechanical}
E.~Glassinger, A.~Lee, R.~Raphael, Electromechanical effects on tether
  formation from lipid membranes: a theoretical analysis, Physical Review E
  72~(4) (2005) 041926.

\bibitem{breneman2009hair}
K.~D. Breneman, W.~E. Brownell, R.~D. Rabbitt, Hair cell bundles: flexoelectric
  motors of the inner ear, PLoS One 4~(4) (2009) e5201.

\bibitem{brownell2001micro}
W.~Brownell, A.~Spector, R.~Raphael, A.~S. Popel, Micro-and nanomechanics of
  the cochlear outer hair cell, Annual review of biomedical engineering 3~(1)
  (2001) 169--194.

\bibitem{deng2019collusion}
Q.~Deng, F.~Ahmadpoor, W.~E. Brownell, P.~Sharma, The collusion of
  flexoelectricity and hopf bifurcation in the hearing mechanism, Journal of
  the Mechanics and Physics of Solids 130 (2019) 245--261.

\bibitem{amit2005field}
D.~J. Amit, V.~Martin-Mayor, Field theory, the renormalization group, and
  critical phenomena: graphs to computers, World Scientific Publishing Company,
  2005.

\bibitem{goldenfeld2018lectures}
N.~Goldenfeld, Lectures on phase transitions and the renormalization group, CRC
  Press, 2018.

\bibitem{kleinert1989gauge}
H.~Kleinert, Gauge fields in condensed matter, Vol.~2, World Scientific
  Singapore, 1989.

\bibitem{kleinert2009path}
H.~Kleinert, Path integrals in quantum mechanics, statistics, polymer physics,
  and financial markets, World scientific, 2009.

\bibitem{bivas1981flexoelectric}
I.~Bivas, A.~G. Petrov, Flexoelectric and steric interactions between two
  bilayer lipid membranes resulting from their curvature fluctuations, Journal
  of theoretical biology 88~(3) (1981) 459--483.

\bibitem{Helfrich1973}
W.~Helfrich, Elastic properties of lipid bilayers: theory and possible
  experiments, Zeitschrift f{\"u}r Naturforschung C 28~(11-12) (1973) 693--703.

\bibitem{Canham1970}
P.~B. Canham, The minimum energy of bending as a possible explanation of the
  biconcave shape of the human red blood cell, Journal of theoretical biology
  26~(1) (1970) 61--81.

\bibitem{abbena2017modern}
E.~Abbena, S.~Salamon, A.~Gray, Modern differential geometry of curves and
  surfaces with Mathematica, CRC press, 2017.

\bibitem{ahmadpoor2017thermal}
F.~Ahmadpoor, P.~Wang, R.~Huang, P.~Sharma, Thermal fluctuations and effective
  bending stiffness of elastic thin sheets and graphene: A nonlinear analysis,
  Journal of the Mechanics and Physics of Solids 107 (2017) 294--319.

\bibitem{steigmann2009concise}
D.~J. Steigmann, A concise derivation of membrane theory from three-dimensional
  nonlinear elasticity, Journal of Elasticity 97~(1) (2009) 97--101.

\bibitem{roohbakhshan2016projection}
F.~Roohbakhshan, T.~X. Duong, R.~A. Sauer, A projection method to extract
  biological membrane models from 3d material models, Journal of the Mechanical
  Behavior of Biomedical Materials 58 (2016) 90--104.

\bibitem{steigmann2013well}
D.~J. Steigmann, A well-posed finite-strain model for thin elastic sheets with
  bending stiffness, Mathematics and Mechanics of Solids 18~(1) (2013)
  103--112.

\bibitem{barham2012magnetoelasticity}
M.~Barham, D.~Steigmann, D.~White, Magnetoelasticity of highly deformable thin
  films: theory and simulation, International Journal of Non-Linear Mechanics
  47~(2) (2012) 185--196.

\bibitem{steigmann2018mechanics}
D.~J. Steigmann, Mechanics and physics of lipid bilayers, in: The role of
  mechanics in the study of lipid bilayers, Springer, 2018, pp. 1--61.

\bibitem{edmiston2011analysis}
J.~Edmiston, D.~Steigmann, Analysis of nonlinear electrostatic membranes, in:
  Mechanics and electrodynamics of magneto-and electro-elastic materials,
  Springer, 2011, pp. 153--180.

\bibitem{ogden2011mechanics}
R.~Ogden, D.~Steigmann, Mechanics and electrodynamics of magneto-and
  electro-elastic materials, Vol. 527, Springer Science \& Business Media,
  2011.

\bibitem{deseri2008derivation}
L.~Deseri, M.~D. Piccioni, G.~Zurlo, Derivation of a new free energy for
  biological membranes, Continuum Mechanics and Thermodynamics 20~(5) (2008)
  255.

\bibitem{mohammadi2014theory}
P.~Mohammadi, L.~Liu, P.~Sharma, A theory of flexoelectric membranes and
  effective properties of heterogeneous membranes, Journal of Applied Mechanics
  81~(1) (2014).

\bibitem{liu2013flexoelectricity}
L.~Liu, P.~Sharma, Flexoelectricity and thermal fluctuations of lipid bilayer
  membranes: Renormalization of flexoelectric, dielectric, and elastic
  properties, Physical Review E 87~(3) (2013) 032715.

\bibitem{grasinger2020statistical}
M.~Grasinger, K.~Dayal, Statistical mechanical analysis of the
  electromechanical coupling in an electrically-responsive polymer chain, Soft
  Matter 16~(27) (2020) 6265--6284.

\bibitem{kleinert1999fluctuation}
H.~Kleinert, Fluctuation pressure of membrane between walls, Physics Letters A
  257~(5) (1999) 269--274.

\bibitem{liu2013energy}
L.~Liu, On energy formulations of electrostatics for continuum media, Journal
  of the Mechanics and Physics of Solids 61~(4) (2013) 968--990.

\bibitem{duttmann2009variational}
M.~D{\"u}ttmann, Variational methods in disorder problems, Ph.D. thesis, PhD
  thesis, Doctoral dissertation, Diploma Thesis), Department of Physics~…
  (2009).

\bibitem{deng2014flexoelectricity}
Q.~Deng, L.~Liu, P.~Sharma, Flexoelectricity in soft materials and biological
  membranes, Journal of the Mechanics and Physics of Solids 62 (2014) 209--227.

\bibitem{ahmadpoor2015thermal}
F.~Ahmadpoor, L.~Liu, P.~Sharma, Thermal fluctuations and the minimum
  electrical field that can be detected by a biological membrane, Journal of
  the Mechanics and Physics of Solids 78 (2015) 110--122.

\bibitem{kleinert1998pressure}
H.~Kleinert, Pressure of membrane between walls, arXiv preprint
  cond-mat/9811308 (1998).

\bibitem{mutz1989unbinding}
M.~Mutz, W.~Helfrich, Unbinding transition of a biological model membrane,
  Physical review letters 62~(24) (1989) 2881.

\end{thebibliography}
\biboptions{sort&compress}

\end{document}